\documentclass[12pt]{article}
\usepackage{graphicx,subfigure}
\usepackage{epsfig}
\usepackage{mathrsfs}
\usepackage{amsmath}
\usepackage{amsfonts}
\usepackage{amssymb}
\usepackage[usenames]{color}
\usepackage[a4paper,left=2.5cm,right=2.5cm,top=2.5cm,bottom=2.5cm]{geometry}

\newcommand{\dd}{\textrm{d}}
\newcommand{\ee}{\mathrm{e}}
\DeclareMathOperator{\waveop}{\square}

\numberwithin{equation}{section}

\title{\LARGE \bf CFTs in rotating black hole backgrounds}

\author{Pau Figueras and Saran Tunyasuvunakool}

\date{}

\begin{document}

\maketitle

\thispagestyle{empty}

\begin{center}
DAMTP, Centre for Mathematical Sciences, Wilberforce Road, Cambridge CB3 0WA, U.K.
\\
\vskip .5cm
\texttt{p.figueras@damtp.cam.ac.uk, s.tunyasuvunakool@damtp.cam.ac.uk}
\end{center}

\vskip 1cm
\begin{abstract}
We use AdS/CFT to construct the gravitational dual of a 5D CFT in the background of a non-extremal rotating black hole. Our boundary conditions are such that the vacuum state of the dual CFT corresponds to the Unruh state. We extract the expectation value of the stress tensor of the dual CFT  using holographic renormalisation and show that it is stationary and regular on both the future and the past event horizons.  The energy density of the CFT is found to be negative everywhere in our domain and we argue that this can be understood as a vacuum polarisation effect. We construct the solutions by numerically solving the elliptic Einstein--DeTurck equation for stationary Lorentzian spacetimes with Killing horizons. 
\end{abstract}

\newpage
\setcounter{page}{1}

\section{Introduction}
\label{sec:intro}

Arguably, the formulation of the AdS/CFT correspondence \cite{Maldacena:1997re,Gubser:1998bc,Witten:1998qj} has been the most important advance in theoretical physics in recent years since it provides a definition of quantum gravity in asymptotically anti-de Sitter (AdS) spacetimes in terms of the dual conformal field theory (CFT). Therefore, this correspondence provides a mean to address fundamental questions in quantum gravity in a well-defined setting. One of the most remarkable effects in quantum gravity is Hawking's realisation that black holes evaporate \cite{Hawking:1974sw}. Hawking's result motivated the study of quantum fields in black hole backgrounds (see \textit{e.g.,} \cite{Birrell:1982ix} for a detailed discussion) but in most of these calculations the fields are assumed to be non-interacting. Adding interactions is very difficult in this approach.  On the other hand, in AdS/CFT we can study strongly interacting quantum fields in curved black hole backgrounds by constructing classical solutions to the bulk Einstein equations with a negative cosmological constant. 

The authors of \cite{Hubeny:2009ru,Hubeny:2009kz,Hubeny:2009rc} have pioneered the study of Hawking radiation in large $N_c$ strongly-coupled field theories which admit gravity duals. As these references pointed out,  the vacuum state in which the CFT is defined is specified by the boundary conditions that the geometry satisfies deep in the IR. Then, in principle, the AdS/CFT dictionary allows one  to extract all physical quantities of the CFT in that particular black hole background in the chosen  vacuum state.

Using the numerical methods developed in \cite{Headrick:2009pv}, Ref. \cite{Figueras:2011va} constructed an Einstein metric in 5D with a negative cosmological constant  such that the boundary geometry is the 4D asymptotically flat Schwarzschild solution. In the IR the geometry reduces to the Poincar\'e horizon of AdS$_5$ and therefore, by the AdS/CFT correspondence, this gravitational solution should describe the large $N_c$, strong coupling limit of $\mathcal N=4$ super Yang-Mills (SYM) on the background of the Schwarzschild black hole in the Unruh state \cite{Figueras:2011va}.\footnote{This reference did not distinguish between the Unruh and Boulware vacua  but \cite{Santos:2012he} has  pointed out that the Boulware state requires a minimum energy configuration and hence an extremal horizon. } This solution allows one to extract the leading $O(N_c^2)$ piece of the expectation value of the quantum stress tensor of the dual field theory in this particular vacuum state. The results of \cite{Figueras:2011va} showed that, remarkably, $\langle T_{ij}\rangle$ is static,  regular on both the future and the past event horizons, and there is no energy flux at infinity.  These results contradict the free field theory expectations (see \textit{e.g.,} \cite{Kay:1988mu}), showing that strong interactions can significantly alter the behaviour of quantum fields, especially in curved spaces. Using the construction of \cite{Figueras:2011va}, Ref. \cite{Figueras:2011gd} subsequently addressed a very closely related problem, namely the construction of a static black hole localised on the brane in the single infinite Randall--Sundrum II (RS2) braneworld model \cite{Randall:1999ee,Randall:1999vf}. In this paper, static braneworld black holes of both small and large radii relative to the radius of the parent AdS space were constructed. These results contradicted the non-existence conjecture of \cite{Tanaka:2002rb,Emparan:2002px,Emparan:2002jp}, which is based on free field theory intuition.\footnote{Refs. \cite{Fitzpatrick:2006cd,Gregory:2008br} had remarked that free field theory intuition might not hold.}

The description of CFTs in black hole backgrounds in the Hartle--Hawking state in terms of their gravity duals requires different boundary conditions in the IR. In the Hartle--Hawking state, the stress tensor of the plasma at infinity should approach that of a thermal fluid in equilibrium with the boundary black hole.  From the bulk point of view, deep in IR the geometry should asymptote to the planar black hole. Ref. \cite{Hubeny:2009ru} further predicted the existence of two families of solutions depending on whether the bulk horizon has a single connected component (black funnels) or two disconnected components (black droplets). The black funnels describe a situation in which the plasma couples strongly to the boundary black hole, resulting in an efficient exchange of heat between them. On the other hand, the black droplets describe the weakly coupled situation. These two phases should be connected through a Gregory--Laflamme type transition.   The black funnels have been subsequently constructed in \cite{Santos:2012he} (and their properties confirmed) whilst the black droplets remain elusive so far (see \cite{Hubeny:2009ru,Hubeny:2009kz,Hubeny:2009rc,Santos:2012he} for more details).

Black funnels have motivated further studies of heat flow in AdS/CFT. Ref. \cite{Fischetti:2012ps} has shown that rotating BTZ black holes in AdS$_3$ can be reinterpreted, by means of a change of conformal frame,  as describing heat flow in the CFT. However, the conformal symmetry in two dimensions implies that the left and right temperatures of the CFT must be constants, and the bulk horizons are Killing. On the other hand, \cite{Fischetti:2012vt} more recently considered a stationary funnel solution in global AdS$_4$ consisting of a single bulk horizon connecting two boundary black holes. The latter have different temperatures which implies that there is a net transport of heat along the bulk horizon, which cannot be  Killing.\footnote{Black holes with non-Killing horizons describing stationary plasma quenches were constructed in \cite{Figueras:2012rb}. }

In this paper we generalise the solutions of \cite{Figueras:2011va} by adding rotation to the boundary black hole. Whilst in principle the Hartle--Hawking state cannot exist in rotating black hole backgrounds with Killing horizons due to issues with superradiance \cite{Kay:1988mu},  we are not aware of any obstructions for the Unruh state. There is an extensive literature on studying \textit{free} quantum fields in static black hole backgrounds (see \cite{Birrell:1982ix} and references therein) but we are not aware of any results in rotating black hole backgrounds in four or higher dimensions. In this paper we remedy this and by using AdS/CFT we compute, for the first time, the expectation value of the stress tensor of a strongly interacting 5D quantum field theory (and in particular a CFT) in a rotating black hole background. Our boundary conditions are such that the vacuum state of the CFT is the Unruh vacuum.

For technical reasons that will be explained subsequently, we construct the gravitational dual of a 5D CFT in the background of the  5D Myers--Perry black hole with equal angular momenta \cite{Myers:1986un}. However,  our construction should capture most of the qualitative physics of the Unruh state in rotating black hole backgrounds for strongly coupled CFTs that admit gravity duals. Another motivation for our work comes form the recent interest in studying 5D superconformal field theories (SCFTs). In general, five dimensional gauge theories are non-renormalisable and therefore do not exist as microscopic theories. However, there are a few notable exceptions. Indeed, for certain choices of gauge group and matter content there exist certain supersymmetric gauge theories in 5D that arise as superconformal fixed points of the renormalisation group \cite{Seiberg:1996bd,Morrison:1996xf,Intriligator:1997pq}. One example of these  gauge theories that is relevant for this paper is 5D super Yang-Mills with gauge group USp$(2N_c)$ and matter consisting of a hypermultiplet in the antisymmetric representation of the gauge group and $N_f$ hypermultiplets in the fundamental representation. This theory arises as the low energy worldvolume theory on $N_c$ D4 branes probing an $O8$ plane with $N_f$ D8 branes on top of it \cite{Seiberg:1996bd}. The theory admits a supergravity dual \cite{Intriligator:1997pq,Brandhuber:1999np} in 10D type IIA massive supergravity. Ref. \cite{Jafferis:2012iv} computed the free energy of this field theory on a $S^5$ using localisation techniques and showed that it scales like $N_c^{\frac{5}{2}}$. Moreover,  the field theory calculation agrees with the supergravity result, which uses the calculation of entanglement entropy \cite{Ryu:2006bv}, generalised to  gravity duals with a non-trivial dilaton \cite{Klebanov:2007ws}.  More recently, Refs. \cite{Bergman:2012kr,Bergman:2012qh} generalised these theories to quiver-type gauge theories and studied their supergravity duals.

The rest of this paper is organised as follows. In \S\ref{sec:setup} we explain our setup and provide the details of the numerical construction of our solutions. In \S\ref{sec:results} we present our results. In \S\ref{subsec:geometry} we consider some geometrical aspects of our solutions and the holographic stress tensor is discussed in  \S\ref{subsec:stresstensor}. Section  \S\ref{sec:conclusions} is devoted to the conclusions and directions for future research. Further details of our solutions as well as convergence tests are relegated to Appendices \ref{sec:pressures} and \ref{sec:convergence} respectively. Appendix \ref{sec:linear} contains the generalisation of the calculation of \cite{Garriga:1999yh} of  the linearised gravitational field on the brane in the RSII model in an arbitrary number of dimensions. 

\textit{Note added}: While our work was nearing completion, we became aware of \cite{Fischetti:2013hja}, which has appeared simultaneously with our paper on the arXiv.  \cite{Fischetti:2013hja} also addresses the construction of the gravitational dual of 5D CFTs in the background of the 5D Myers--Perry black hole with equal angular momenta. 

\section{Rotating boundary black holes in Poincar\'e AdS}
\label{sec:setup}

In this section we describe our construction of rotating black holes in Poincar\'e AdS. More precisely, we will consider Einstein metrics in 5D with a negative cosmological constant such that the boundary metric is conformal to the 5D Myers--Perry  black hole \cite{Myers:1986un}  with equal angular momenta. In the region of the spacetime which is far from the horizon, our geometry approaches the Poincar\'e horizon of AdS$_6$. Therefore, by the gauge gravity duality, our solution is dual to a CFT on a rotating black hole background.  As discussed in \cite{Figueras:2011va,Santos:2012he}, our boundary conditions in the IR should imply that the dual CFT is in the Unruh vacuum state. 

Since the boundary horizon is rotating (and Killing), by continuity the bulk horizon will have the same angular velocity as the boundary black hole. Therefore, our solutions are the first examples of rotating horizons in the Poincar\'e patch of AdS.\footnote{The planar limit of a rotating black holes in global AdS yields a boosted black brane.}  Moreover, we shall restrict ourselves to the case of \textit{non-extremal} black holes, and the temperatures of the boundary and bulk horizons are the same. 

The reason why we consider the equal angular momenta Myers--Perry black hole as our boundary geometry is a technical one: such a spacetime is of co-homogeneity one, and therefore the full spacetime metric will depend only on two coordinates (see \S\ref{subsec:ansatz} below). The problem then reduces to that of solving partial differential equations (PDEs) in two variables. We believe that our solution should capture the essential physics of quantum field theories on rotating black hole backgrounds. In some respects,  the properties of the 5D Myers--Perry solution with equal angular momenta resemble those of the 4D Kerr solution (\textit{e.g.}, existence of a smooth extremal limit, non-existence of an ultraspinning regime \cite{Dias:2010eu}), and therefore we expect that the physics of $\mathcal N=4$ SYM on the Kerr black hole background should be qualitatively similar to what is described in this paper.   Nevertheless, it would still be interesting to study the latter. In this other case, the problem is of co-homogeneity 3 (but still tractable with our techniques) and we will leave it for future work. 

We want to construct a solution to the Einstein equations in 5D with a negative cosmological constant obeying the boundary conditions outlined above. To do so, we will use the method of \cite{Headrick:2009pv,Adam:2011dn} (see \cite{Wiseman:2011by} for a review) and instead of solving the Einstein equations, we will solve the so called Einstein--DeTurck equations (also referred as the `Harmonic' Einstein equations),
\begin{equation}
R^H_{ab}\equiv R_{ab} + \frac{5}{\ell^2}\,g_{ab} + \nabla_{(a}\xi_{b)} = 0 \,, \qquad \xi^a= g^{bc}(\Gamma^a_{\phantom a bc} - \bar\Gamma^a_{\phantom a bc})\,,
\label{eqn:deturck}
\end{equation}
where $\ell$ denotes the radius of AdS$_6$, $\Gamma^a_{\phantom a bc}$ is the Levi-Civita connection  associated to the spacetime metric $g$ that we want to determine. For simplicity,  we take $\bar\Gamma^a_{\phantom a bc}$ to be the Levi-Civita connection associated to some chosen reference metric $\bar g$ on the same spacetime manifold $\mathcal M$. Solutions to \eqref{eqn:deturck} are Einstein iff $\xi^a=0$ everywhere on $\mathcal M$. To achieve this,  we shall impose boundary conditions (see \S\ref{subsubsec:BCs}) which are compatible with $\xi^a|_{\partial\mathcal M} = 0$ \cite{Headrick:2009pv,Figueras:2011va,Adam:2011dn} and \emph{a posteriori} we find that indeed $\xi^a=0$ everywhere in our domain (within our numerical accuracy).

\subsection{Equal angular momenta 5D Myers--Perry black hole}

In this subsection, we begin with a review of the geometry of the 5D Myers--Perry black hole with equal angular momenta. We then introduce new coordinates which will be useful for our subsequent numerical construction of the gravitational duals of CFTs in rotating black hole backgrounds. 

The 5D Myers--Perry black hole with equal angular momenta is described by the following line element \cite{Myers:1986un}:
\begin{equation}
\dd s^2_{\textrm{MP}_5} = \phantom{} - \dd t^2 + \left[ \frac{R^2 (R^2 + a^2)}{(R^2 + a^2)^2 - \mu R^2} \right] \dd R^2 + \left[ \frac{\mu}{R^2 + a^2} \right] \left( \dd t + \frac{a}{2} \sigma^3 \right)^2 + \left( R^2 + a^2 \right) \dd \Omega_{(3)}^2 \, ,
\label{eqn:MP5tmp}
\end{equation}
where $\mu$ and $a$ are the mass and angular momentum parameters respectively,  and
\begin{equation}
\begin{aligned}
\sigma^1 &= - \sin \psi \dd \theta + \sin \theta \cos \psi \dd \phi \\
\sigma^2 &= \cos \psi \dd \theta + \sin \theta \sin \psi \dd \phi \\
\sigma^3 &= \cos \theta \dd \phi + \dd \psi
\end{aligned}
\end{equation}
are the left-invariant one-forms on SU$(2)$. Here $\dd \Omega_{(3)}^2 = \frac{1}{4} \left( (\sigma^1)^2 + (\sigma^2)^2 + (\sigma^3)^2 \right) $ is the round metric on the unit 3-sphere. The event horizon is located at $R=R_H$ where $R_H$ is the largest real root of the equation $(R_H^2 + a^2)^2 - \mu R_H^2 = 0$. In the following it will be useful to express the mass parameter $\mu$ in terms of $R_H$ and $a$:
\begin{equation}
\mu = \frac{(R_H^2 + a^2)^2}{R_H^2}\,.
\end{equation}
It is then straightforward to find a convenient compact radial coordinate by defining
\begin{equation}
R^2 + a^2 = \frac{R_H^2 + a^2}{(1-r^2)^2}\,,
\end{equation}
so that $r=0$ corresponds to the event horizon (or, more precisely, the bifurcation surface) and $r\to 1$ corresponds to spacelike infinity. In terms of the new radial coordinate $r$, we can rewrite \eqref{eqn:MP5tmp} as
\begin{equation}
\dd s^2_{\textrm{MP}_5} = \phantom{} - r^2 T_0(r) \, \dd t^2 + \frac{4 R_0(r)}{\left( 1-r^2 \right)^4} \, \dd r^2 + \frac{B_0(r)}{4 \left( 1-r^2 \right)^2} \, \left( \sigma^3 - \Omega_0(r) \dd t \right)^2 + \frac{S_0}{4 \left( 1-r^2 \right)^2} \, \dd \Omega_{(2)}^2\,,
\label{eqn:MP5}
\end{equation}
where
\begin{equation}
\begin{aligned}
R_0(r) &= \frac{R_H^2 \left( R_H^2 + a^2 \right)}{ \left( 2 - r^2 \right) \left( R_H^2 - a^2 (1 - r^2)^2 \right)} \\
S_0(r) &= R_H^2 + a^2\\
T_0(r) &= \frac{\left( 2 - r^2 \right) \left( R_H^2 - a^2 (1-r^2)^2 \right)}{R_H^2 + a^2 \left( 1-r^2 \right)^4} \\
B_0(r) &= \frac{\left( R_H^2 + a^2 \right) \left( R_H^2 + a^2 (1-r^2)^4 \right)}{R_H^2} \\
\Omega_0(r) &= \phantom{} - \frac{2 a \left( 1-r^2 \right)^4}{R_H^2 + a^2 \left(1-r^2\right)^4},
\end{aligned}
\label{eqn:funcB}
\end{equation}
and $\dd \Omega_{(2)}^2 = (\sigma^1)^2 + (\sigma^2)^2$ is the metric on the round unit 2-sphere. Note that when written in this form, the metric exhibits the structure of a U$(1)$ fibration over an $S^2$ base.  As we shall see in \S\ref{subsec:ansatz}, we can consistently construct an extension of this metric into the bulk of AdS$_6$ whilst preserving this structure.  In these coordinates, the null generator of the future event horizon is given by, 
\begin{equation}
k = \frac{\partial}{\partial t} + \Omega_H \frac{\partial}{\partial \psi},
\end{equation}
and the angular velocity and surface gravity of the horizon are 
\begin{equation}
\Omega_H = \Omega_0(0) = \phantom{}  \frac{2a}{R_H^2 + a^2} \, ,\qquad 
\kappa^2 = \frac{T_0(0)}{4R_0(0)} = \frac{\left(R_H^2 - a^2\right)^2}{R_H^2 \left( R_H^2 + a^2 \right)^2} \,.
\label{eqn:omegakappa}
\end{equation}

The form of the metric \eqref{eqn:MP5} is particularly useful because the boundary conditions at the horizon that the various functions satisfy become transparent (see \S\ref{subsec:ansatz}). In fact, any rotating black hole solution with a Killing horizon and with an orthogonally transitive isometry group can be cast in an analogous form, so the boundary conditions that we discuss below are general. Alternatively, we could have chosen to work in co-rotating coordinates as in \cite{Adam:2011dn} at the expense of losing the manifest asymptotic flatness of the metric. This could potentially be problematic when extending the metric into the AdS bulk if we want to require that the spacetime asymptotes the Poincar\'e horizon of AdS$_6$ in the region far from the horizon.

\subsection{Metric ansatz and boundary conditions}
\label{subsec:ansatz}

Let us consider first the metric on the Poincar\'e patch of AdS$_6$:
\begin{equation}
\dd s^2_{\textrm{AdS}_6} = \frac{\ell^2}{z^2}(\dd z^2 - \dd t^2 + \dd R^2 + R^2\dd \Omega_{(3)}^2 )\,.
\label{eqn:PAdS}
\end{equation}
For our construction it is convenient to introduce new coordinates which are adapted to the Poincar\'e horizon of AdS. To do so, we generalise the coordinate change introduced in \cite{Figueras:2011va} and define new coordinates $(r,x)$ as
\begin{equation}
z = \frac{ \left( 1 - x^2 \right) \sqrt{ R_H^2 + a^2 } }{1 - r^2} \,,\qquad R = \frac{x \, \sqrt{ \left( 2-x^2 \right) \left( R_H^2 + a^2 \right) } }{1 - r^2} \, ,
\end{equation}
in terms of which \eqref{eqn:PAdS} becomes 
\begin{equation}
\dd s^2_{\textrm{AdS}_6} =  \frac{\ell^2 \left( 1 - r^2 \right)^2}{\left(1 - x^2\right)^2} \left[ \frac{4 \, \dd x^2}{\left(2-x^2\right) \left(1 - r^2\right)^2}  - \frac{\dd t^2 }{R_H^2 + a^2} + \frac{4 r^2 \, \dd r^2}{\left( 1-r^2 \right)^4} + \frac{x^2 \left(2 - x^2 \right)}{\left( 1 - r^2 \right)^2} \, \dd \Omega_{(3)}^2 \right] \, .
\end{equation}
In these coordinates the conformal boundary of AdS is in the asymptotic region $x\to 1$, and $x=0$ is the fixed point set of the SO$(4)$ symmetry. These coordinates are adapted to the Poincar\'e horizon of AdS in the sense that the latter lies at $r=1$ and $x<1$. The point $x\to 1$ and $r\to 1$ corresponds to spacelike infinity at the boundary.

We want to construct a 6-dimensional asymptotically locally AdS spacetime  such that its boundary metric is in the same conformal class as the 5D Myers--Perry black hole, and that approaches the Poincar\'e horizon of AdS far from the conformal boundary.  The isometry group of the boundary metric is $\mathbb R_t\times$SU$(2)\times$U$(1)$. It has been shown in 4D that regular and \textit{static} asymptotically locally AdS metrics inherit the isometry group of the boundary metric \cite{Anderson:2002xb}.  We shall make the reasonable assumption that in higher dimensions and in the presence of rotation the same result holds true. Therefore, our full spacetime metric should have the full $\mathbb R_t\times$SU$(2)\times$U$(1)$ isometry group of the boundary metric. Then, the most general metric (\textit{i.e.,} closed under diffeomorphisms that preserve the symmetries) that satisfies our  symmetry assumptions is given by
\begin{align}
\dd s^2 = \frac{\left( 1 - r^2 \right)^2}{\left(1 - x^2\right)^2} &\left[ \frac{4 \, X(x,r)}{\left(2 - x^2\right) \left(1 - r^2\right)^2} \, \dd x^2 + \frac{2 \, x \, r \left(R_H^2 + a^2\right) Y(x,r)}{\left( 1 - r^2 \right)^3} \, \dd x \,\dd r \right. \nonumber\\
&\left. \quad \phantom{} - r^2 \, T_0(r) \, T(x,r) \, \dd t^2 + \frac{x^2 \left( 2-x^2 \right) B_0(r) \, B(x,r)}{4 \left( 1-r^2 \right)^2} \left( \sigma_3 - \Omega(x,r) \dd t \right)^2 \right.  \nonumber \\
&\left. \quad \phantom{}+ \frac{x^2 \left( 2-x^2 \right) S_0(r) \, S(x,r)}{4 \left(1-r^2\right)^2} \, \dd \Omega_{(2)}^2 
+ \frac{4 \, R_0(r) \, R(x,r)}{\left( 1-r^2 \right)^4} \, \dd r^2\right] \, , \label{eqn:ansatz}
\end{align}
where $\{X(x,r),Y(x,r),T(x,r),B(x,r),\Omega(x,r),R(x,r),S(x,r)\}$ are the unknown functions to be solved for,  and $\{T_0(r), R_0(r), B_0(r),S_0(r)\}$ are the Myers--Perry functions given in \eqref{eqn:funcB}. As above, the coordinate ranges are $0\leq x < 1$ and $0\leq r < 1$ but we will effectively treat the asymptotic regions $x=1$ (conformal boundary of AdS) and $r=1$ (Poincar\'e horizon) as boundaries and work on the square domain $0\leq x \leq 1$ and $0\leq r \leq 1$. On this domain we shall assume that all our unknown functions are smooth and bounded.  Note that because we have chosen coordinates which make the isometries of the metric manifest (and hence reducing the computational cost), $r=0$ (event horizon) and $x=0$ (fixed point set of the SO$(3)$ symmetry) appear as boundaries of our domain.  In fact, we can identify the square $\{0\leq r \leq 1,\, 0\leq x \leq 1\}$ with the orbit space $\mathcal O \cong\mathcal M/[\mathbb R_t\times$SU$(2)\times$U$(1)]$, which has the natural structure of a manifold with boundaries and corners.  However, as discussed in \cite{Figueras:2011va,Adam:2011dn},  some of these boundaries are in fact only fictitious boundaries after suitable smoothness conditions on the metric functions are imposed there.

\subsubsection{Boundary conditions} 
\label{subsubsec:BCs}
Refs. \cite{Figueras:2011va,Adam:2011dn} discuss in generality the behaviour of the metric on different kinds of boundaries that are relevant for our problem. In the following we shall specialise these general results to our case. 

Comparing \eqref{eqn:ansatz} with \eqref{eqn:MP5} we see that at $x=1$ we have to impose the following Dirichlet boundary conditions:
\begin{equation}
\begin{aligned}
&X(1,r) = \ell^2\,, \quad Y(1,r) = 0\,, \quad \Omega(1,r) = \Omega_0(r) \, ,\\
&T(1,r) = R(1,r) = B(1,r) = S(1,r) =\frac{\ell^2}{R_H^2 + a^2}\,. 
\end{aligned}
\end{equation}
This choice ensures that as $x\to 1$ the metric becomes
\begin{equation}
\dd s^2 \sim \frac{\ell^2 \left( 1 - r^2 \right)^2}{\left(1 - x^2\right)^2} \left[ \frac{4}{\left( 2-x^2 \right) \left( 1-r^2 \right)^2} \, \dd x^2 + \frac{1}{\left( R_H^2 + a^2 \right)} \, \dd s^2_{\textrm{MP}_5} \right] \, ,
\end{equation}
which shows that the boundary metric lies in the same conformal class as the 5D Myers--Perry black hole. 
It is worth mentioning that we allow for the constant factor of $\left(R_H^2+a^2\right)^{-1}$ multiplying $\dd s^2_{\textrm{MP}_5}$ so that the boundary conditions at $x=1$ and $r=1$ (see below) are compatible for any choice of $R_H$ and $a$. This is necessary to ensure smoothness of the unknown functions.

Following the general discussion on boundary conditions for extremal horizons in \cite{Figueras:2011va}, since at $r\to 1$ we want our metric to smoothly approach the Poincar\'e horizon of AdS$_6$, we must impose
\begin{align}
&X = \ell^2 + (1-r)\,X_1\,,\quad Y = (1-r)\,Y_1\,,\quad T = \frac{\ell^2}{R_H^2+a^2}+(1-r)\,T_1\,,\nonumber\\
& R = \frac{\ell^2}{R_H^2+a^2} + (1-r)\,R_1 \,,\quad B =  \frac{\ell^2}{R_H^2+a^2} + (1-r)\,B_1\,,\quad S =  \frac{\ell^2}{R_H^2+a^2} + (1-r)\,S_1\,,\nonumber\\
&\Omega = (1-r)\Omega_1 \,,\quad T_1|_{r=1} - R_1|_{r=1} = \textrm{constant} 
\end{align}
where all the functions with the subscript $_1$ are smooth at $r=1$ and $x<1$. Note that $R_0(1)=B_0(1)=S_0(1)=R_H^2+a^2$ and also that $\dd \Omega_{(3)}^2 = \frac{1}{4} \left( \dd \Omega_{(2)}^2 + \left( \sigma^3 \right)^2 \right)$. As shown in \cite{Figueras:2011va}, these boundary conditions are necessary and sufficient for our spacetime to asymptote the Poincar\'e horizon of AdS$_6$. In fact, we shall not impose the condition $T_1|_{r=1} - R_1|_{r=1} = \textrm{constant}$ but we check that it is satisfied on our solutions as consequence of the equations of motion. Hence, we can use this condition as an estimate of the numerical error of our solutions.

At $x=0$ both the $S^2$ of the base space and the U$(1)$ fiber degenerate. To ensure that they do so smoothly, the functions $X,Y,T,R,B,\Omega,S$ must be smooth functions of $x^2$ near $x=0$ and hence they should obey a Neumann boundary condition there. In addition, to avoid the presence of conical singularities at $x=0$ we must require
\begin{equation}
\frac{ B_0(r) B(0,r) }{ X(0,r) } = \frac{ S_0(r)\, S(0,r) }{ X(0,r) } = 1 \quad \textrm{at}\quad x=0\,.
\label{eqn:axisreg}
\end{equation}
Smoothness of the manifold at the horizon requires all metric functions to be smooth in $r^2$ near $r=0$, and hence they obey a Neumann boundary condition there. In addition, as discussed in \cite{Adam:2011dn}, we must ensure that our metric has the same surface gravity and angular velocity as the reference metric (see below for our specific choice of reference metric), which can be achieved imposing
\begin{equation}
T(x,0)=R(x,0)\,,\qquad \Omega(x,0)=\Omega_0(0)\,.
\end{equation} 
Then the surface gravity and angular velocity of our spacetime will be given by \eqref{eqn:omegakappa}.

In order to solve the Einstein--DeTurck equations \eqref{eqn:deturck} we have to specify a reference metric $\bar g$ on the same manifold $\mathcal M$ as the physical spacetime metric $g$ that we are seeking. This is equivalent to choosing a particular gauge. In our case, we have chosen $\bar g$ to be given by \eqref{eqn:ansatz} with 
\begin{equation}
\begin{aligned}
&X=\ell^2\,,\quad Y=0\,,\quad T=R=S=\frac{\ell^2}{R_H^2+a^2}\,,\\
&\Omega = \Omega_0\,, B = \frac{\ell^2}{R_H^2+a^2}\bigg[x^2 + \frac{R_H^2(1-x^2)}{B_0(R_H^2+a^2)}\bigg]\,.
\end{aligned}
\end{equation}

Finally we note that the boundary conditions spelled out above are compatible with $\xi^a|_{\partial\mathcal M}=0$. In the static case and for boundary conditions similar to ours, Ref. \cite{Figueras:2011va} proved (under suitable smoothness assumptions) that all solutions to \eqref{eqn:deturck} obeying these boundary conditions should satisfy  $\xi^a=0$ everywhere on $\mathcal M$ and therefore are solutions to the Einstein equations as well. An analogous result that applies to stationary metrics is not available yet so, given our boundary conditions, we cannot rule out \emph{a priori} the existence of `Ricci solitons' (\textit{i.e.} solutions to \eqref{eqn:deturck} with $\xi^a\neq 0$).  However, because the problem is elliptic, the solutions should be locally unique. Therefore an Einstein metric cannot be arbitrarily close (in the space of solutions) to a Ricci soliton \cite{Headrick:2009pv,Figueras:2011va,Adam:2011dn}.  Therefore, we can always check \emph{a posteriori} for a given solution whether $\xi^a=0$. Needless to say, for the solutions presented in \S\ref{sec:results} we find $\xi^a\to 0$ in the continuum limit.

\subsection{Details of the numerics}

In this subsection we present the details of our numerical solution of \eqref{eqn:deturck} with the boundary conditions specified in the previous subsection. 

To solve \eqref{eqn:deturck} we considered the associated parabolic equation,  namely the  Ricci--DeTurck flow equation, and evolved the metric according to
\begin{equation}
\frac{\partial}{\partial\lambda}\,g_{ab} = -2\,R^H_{ab}\,,
\label{eqn:flow}
\end{equation}
until we reached a fixed point and hence also a solution of \eqref{eqn:deturck}. Here $\lambda$ is the Ricci flow time. In order to carry out the flows we have to provide an initial metric $g_{ab}|_{\lambda=0}$ and for the results presented in \S\ref{sec:results} we chose the initial metric to be equal to the reference metric. The results should not be qualitatively different for other choices of initial data, but we have not investigated this. 

As discussed in \cite{Headrick:2006ti,Headrick:2009pv}, the convergence of the flow towards the fixed point depends on the stability of the fixed point.  The latter is determined by the presence of negative modes in the spectrum of the Lichnerowicz operator computed around the fixed point.\footnote{The linearisation of \eqref{eqn:deturck} around an Einstein metric coincides with the Lichnerowicz operator so in this discussion we are assuming that the fixed points are Einstein metrics. This is indeed the case for the solutions presented in this article. } In our simulations we found that our initial data always converged to the desired Einstein metric with no fine-tuning,\footnote{See \cite{Headrick:2009pv} for a detailed explanation of how to use Ricci flow as a method to construct Einstein metrics possessing one negative mode.} indicating that our solutions are stable fixed points of the Ricci--DeTurck flow and hence do not posses negative modes.  Even though the 5D Myers--Perry black hole has a negative mode \cite{Dias:2010eu} (completely analogous to the celebrated negative mode of 4D Schwarzschild \cite{Gross:1982cv} or Kerr \cite{Monteiro:2009ke}), it seems that pinning the metric to the boundary of AdS projects out this negative mode. Ref.  \cite{Figueras:2011va} observed the same phenomenon for the 5D bulk case in which the boundary metric is simply 4D Schwarzschild. It would be interesting to have a more detailed understanding of why and how this happens.

In AdS/CFT only the conformal class of boundary metrics matters and therefore we can choose any representative within the  class without loss of generality. For the problem at hand,  choosing a representative of the conformal class amounts to fixing the overall length scale of the boundary 5D Myers--Perry black hole. In the results presented below, we fixed the mass parameter $\mu$ to some convenient value and varied the angular momentum parameter $a$ to move along the family of solutions. Note that this implies that both the surface gravity and angular velocity vary along this family.

As in \cite{Headrick:2009pv}, there are two main sources of numerical error. The first one is present in any numerical method and it is simply the fact that we discretise the equations \eqref{eqn:flow} according to some scheme and seek a solution to the discrete problem. The latter approximates the continuum solution and, for a consistent discretisation, in the continuum limit the error should decrease at some rate which depends on the discretisation scheme and the degree of differentiability of the solution. The second source of error is related to the fact that it takes infinite Ricci flow time for \eqref{eqn:flow} to reach its fixed point, but in practice we can only evolve \eqref{eqn:flow} for a finite time. We then have to make sure that we are sufficiently close to the fixed point so that the dominant error is the discretisation error. We do so by monitoring $|\ell^2\,R/30 + 1|_{\textrm{max}}$ (which vanishes for an Einstein metric in the continuum limit) along the flow and we only stop the simulation when this quantity is constant within some tolerance. The value of this constant gives an estimate of the numerical error of our solution and it decreases as we increase the spatial resolution; in Appendix \ref{sec:convergence} we present more details about the convergence tests that we have performed.  Of course, the amount of Ricci flow time that it takes to reach the fixed point depends on how close to extremality the solution is and the resolution of the spatial discretisation.

We simulated the Ricci--DeTurck flow using the method of lines. The two spatial dimensions were discretised onto a square grid of Chebyshev points, and spatial derivatives were approximated by spectral differentiation. The flow was then time-evolved using a third-order Runge--Kutta integrator with a fixed step size. We note also that spatial differentiation is equivalent to simple pointwise multiplications in the spectral space, and therefore we can improve the computational speed by using the discrete cosine transform rather than repeated matrix multiplications.

In our calculation of the curvature tensor, we hard-coded the expression for each of the non-zero components in terms of the metric components and their derivatives. Due to the complexity of this problem, a typical component involves a large number of additions and subtractions of floating point quantities. As the metric becomes singular near the horizon, the relative sizes of these summands can become vastly different. At double-precision, we found that this becomes problematic when we get sufficiently close to the fixed point: numerical error starts to dominate and the flow oscillates wildly and does not converge any further. The problem worsens as we increase the number of grid points since the metric components need to be computed closer to the horizon. 

The problem is exacerbated by the fact that, in order to extract the boundary stress tensor, we need to read off the fifth derivative from our numerical data, and for each extra derivative that we require, the numerical results needs to be evaluated more precisely. In some cases, we found that in order to extract the fifth derivative to an acceptable quality, we had to invoke quadruple-precision floating point arithmetic in our numerical computation. Unfortunately, as no current equipment implements this at hardware level, we had to rely on software libraries to emulate higher-precision floating point arithmetic, which can be over 20 times slower than native computation.

To optimise the time that it takes to obtain solutions with acceptable numerical accuracy, we divided the typical run into three phases. We first evolve the Ricci flow equation numerically using the standard 64-bit double-precision floating point arithmetic. Once our error indicators (Ricci scalar and norm of DeTurck vector) start to oscillate, we took the numerical solution at the final Ricci flow time and use it as initial data for a second run, which is carried out using 80-bit ``long double'' precision. Finally, we take our solution from the second run as the initial data for a final run at 128-bit quadruple precision. Typically it takes between one and two days to complete the whole process. In all cases, the data presented in the next section is the result of this final run at quadruple precision.

For the majority of cases, we found that a $36 \times 36$ square Chebyshev grid provided a good balance between accuracy and computational requirements. However, for the slowly-spinning cases we also found that finer grids are required in order to maintain an acceptable accuracy (we chose a $45 \times 45$ grid for the three most slowly-spinning cases).

\section{Results}
\label{sec:results}


In this section we present our results: \S\ref{subsec:geometry} discusses some aspects of the geometry of the solutions and in  \S\ref{subsec:stresstensor} we present the holographic stress tensor together with its interpretation. To construct our solutions we have fixed the mass parameter of the boundary black hole, $\mu$, and then varied the angular momentum parameter $a$ such that the dimensionless surface gravity is $\kappa\,\mu^{\frac{1}{2}}=\frac{n}{16}$ with $n=1,\ldots, 16$. Therefore, for our solutions the horizon of the black hole is always non-extremal.

\subsection{Geometry of the horizon}
\label{subsec:geometry}

\begin{figure}[t]
\begin{center}
\includegraphics[scale=0.455]{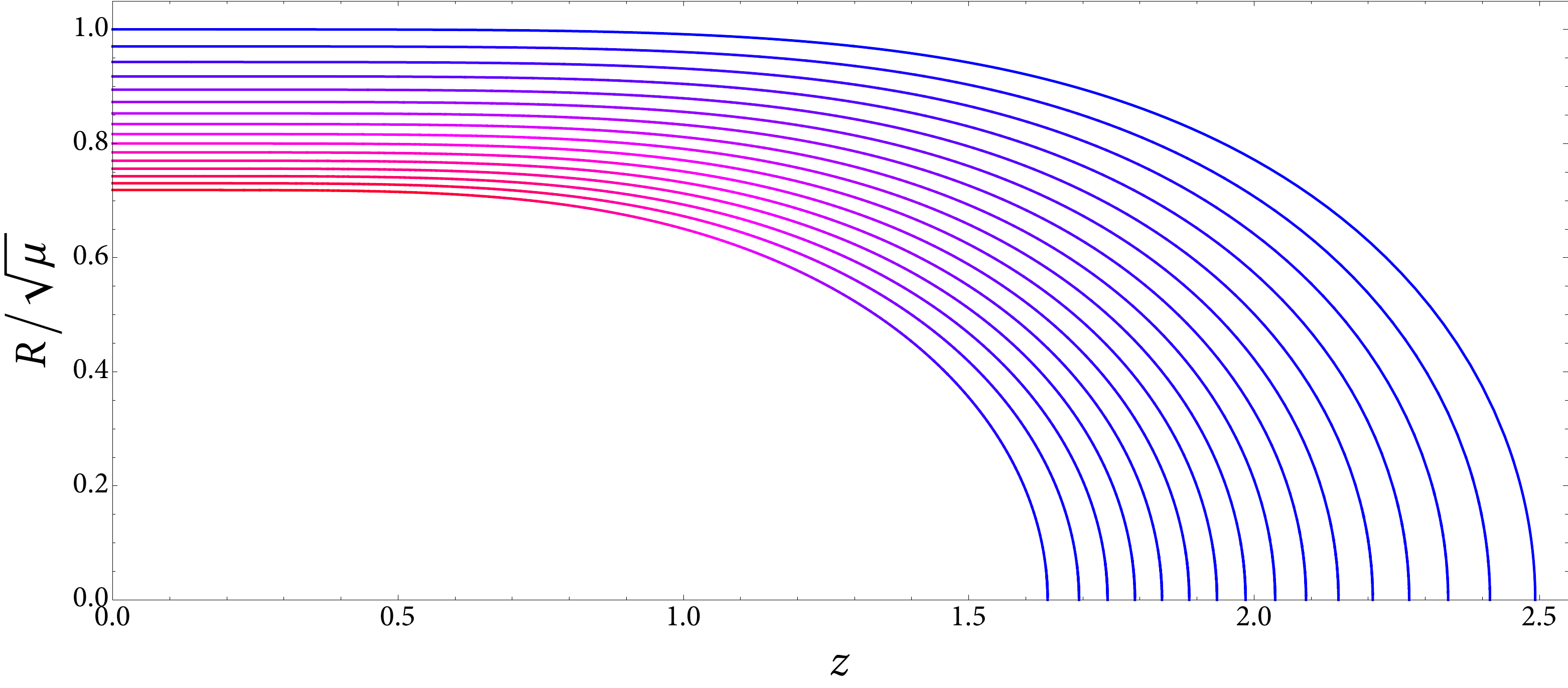}
\hspace{0.25cm}
\includegraphics[scale=0.21]{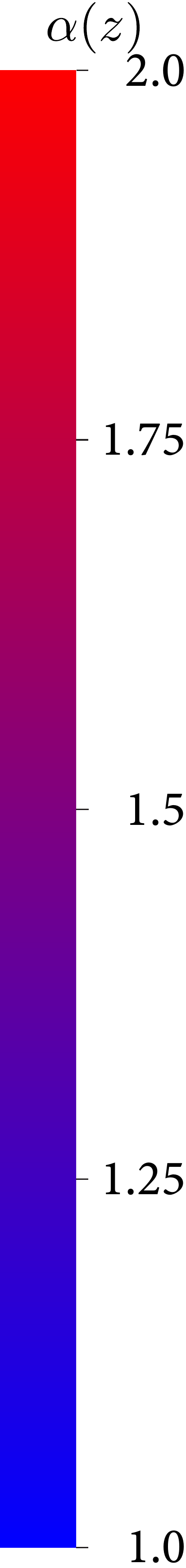}
\end{center}
\caption{Embeddings of the radius $R(z)$ of the horizon $S^2$ into four-dimensional hyperbolic space for fixed mass parameter $\mu$ of the boundary black hole. At the boundary of hyperbolic space, $z=0$, the value of this radius function coincides with the corresponding one in  the 5D Myers--Perry black hole geometry. The colour code depicts the squashing of the horizon $S^3$ according to the relative size of the fibre with the respect to the base $S^2$, which is measured by the function $\alpha(z) = B_0(0)B(z,0)/(S_0(0)S(z,0))$. The blue curve with $R(0)=1$ corresponds to the zero rotation case and $a$ increases moving inwards in this plot. The squashing becomes more severe near the boundary and with increasing angular momentum of the boundary black hole. Absence of conical singularities at the axis of symmetry $R(z_\textrm{max})=0$ implies that there is no squashing there.}
\label{fig:embedding}
\end{figure}

To visualise the horizon geometry of our solutions (\textit{i.e.,} fixed points of \eqref{eqn:flow}) as we vary the angular momentum of the boundary black hole, we consider the following embedding. We take the geometry of the spatial cross-sections of the horizon transverse to the U$(1)$ fiber, 
\begin{equation}
\dd s_H^2 = \frac{\ell^2}{(1-x^2)^2}\bigg[\frac{4\,X(x,0)}{2-x^2}\,\dd x^2 + \frac{x^2(2-x^2)S_0(0)\,S(x,0)}{4}\,\dd \Omega_{(2)}^2\bigg]\,,
\end{equation}
and embed them into 4D hyperbolic space with the same radius $\ell$, $\dd s^2_{\mathbb H_4}=\frac{\ell^2}{z^2}(dz^2 + \dd R^2 + R^2 \,\dd\Omega_{(2)}^2)$, as a curve $R=R(z)$. Then we colour-code this curve according to the relative size of the U$(1)$ fibre with respect to the base $S^2$, which is measured by the function $\alpha(z) = B_0(0)B(x,0)/(S_0(0)S(x,0))$ where the coordinate $x$ should be regarded as being a function of the coordinate $z$ of the ambient hyperbolic space. The freedom in the embedding is fixed by requiring that for the static case the radius of the boundary black hole is $R(0)=1$, which fixes the ADM mass of the boundary black hole for any $a\geq 0$.  The results are depicted in Fig. \ref{fig:embedding}. As this figure shows, the radius of the $S^2$ becomes smaller as angular momentum increases; this is simply the reflection of the fact that the horizon area at fixed mass decreases as the angular momentum increases. Also, note that the extent of the horizon into the bulk decreases as the angular momentum increases. Further,  for a given fixed value of the angular momentum parameter $a>0$,  the squashing of the horizon three-sphere is more severe at the boundary rather than at the interior of the spacetime.  In fact, the regularity condition \eqref{eqn:axisreg} at the axis of symmetry, $x=0$, implies that the the horizon $S^3$ must be locally round there, and therefore there cannot be any squashing.

\begin{figure}[t]
\begin{center}
\includegraphics[scale=0.2]{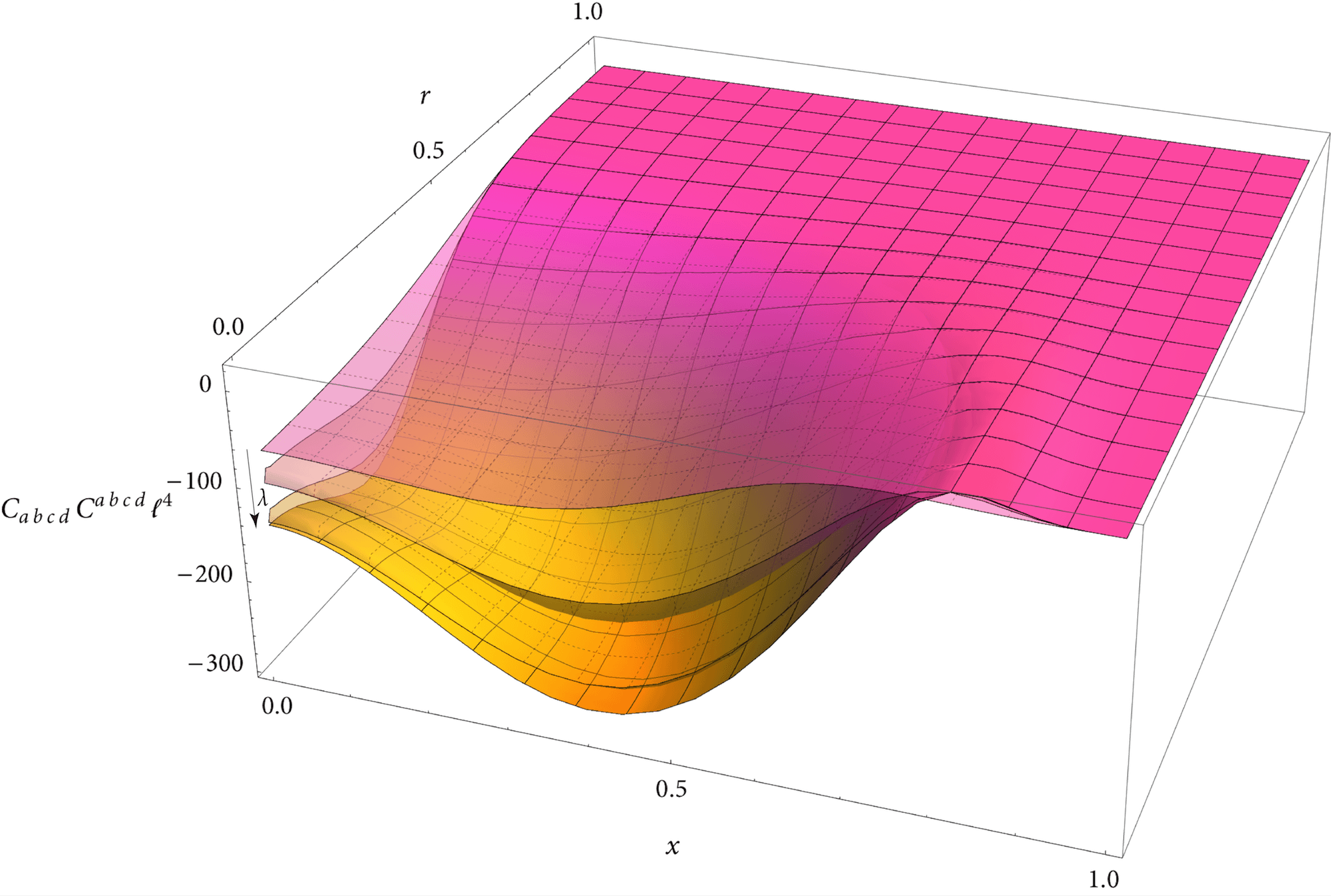}
\end{center}
\caption{$C_{abcd}C^{abcd}\,\ell^4$ along the Ricci flow in the whole domain for the $\kappa\,\mu^{\frac{1}{2}} = 0.15 $ solution. The snapshots correspond to the Ricci flow time $\lambda = 0,\,0.5,\,1$ and the fixed point (from top to bottom in this plot). After Ricci flow time of $\lambda = 1$ the surface becomes almost indistinguishable from that of the actual fixed point. As this plot shows, $C_{abcd}C^{abcd}\,\ell^4$ does not blow up anywhere in the domain along the flow. Also, it vanishes at both the Poincar\'e horizon and at the boundary of AdS, which indicates that our spacetime is asymptotically AdS near these boundaries. }
\label{fig:weyl2}
\end{figure}

In Fig. \ref{fig:weyl2}  we depict the evolution of the curvature invariant $C_{abcd}C^{abcd}\,\ell^4$ along the flow for a typical solution. This is a useful geometric quantity because it gives information of both the curvature of the spacetime and the correctness of our boundary conditions. As this figure shows, $C_{abcd}C^{abcd}\,\ell^4$ never blows up along the flow (and in particular at the fixed point), which suggests that there are no curvature singularities in our domain. In addition,  we see that for any $\lambda$ this curvature invariant vanishes at both the Poincar\'e horizon of AdS$_6$ and at the conformal boundary. This confirms that our spacetime indeed has the correct asymptotics near these two boundaries.

\subsection{Stress tensor}
\label{subsec:stresstensor}

We extract the vacuum expectation value (vev) of the stress tensor of the dual CFT from the bulk gravity solution using the standard holographic renormalisation prescription \cite{deHaro:2000xn}. For a 5D bulk spacetime, the expectation value of the stress of tensor the dual CFT is given by
\begin{equation}
\langle T_{ij} \rangle = \frac{5\,\ell^4}{16\,\pi\,G_6}\,g^{(5)}_{ij}\,,
\label{eqn:dualstress}
\end{equation}
where $g^{(5)}_{ij}$ denotes the 5th order term in the near boundary expansion of the metric in Fefferman--Graham coordinates and $G_6$ is the 5D bulk Newton's constant.  

  Our bulk 5D solutions can be trivially uplifted on half an $S^4$ (see \cite{Cvetic:1999un}) to massive type IIA supergravity in 10D, which arises as the low energy limit of intersections of D4 and D8 branes in type I$'$ string theory.\footnote{Note that there exist other uplifts of Romans' supersymmetric vacuum leading to new supersymmetric AdS$_6$ solutions in type II supergravity  with potentially well-defined CFT duals \cite{Lozano:2012au,Jeong:2013jfc}.} The CFT that arises as the fixed point of this system under RG flow is USp$(2N_c)$ 5D SYM with matter consisting of a hypermultiplet in the antisymmetric representation of the gauge group and $N_f$ hypermultiplets in the fundamental \cite{Brandhuber:1999np}. This is  the CFT dual to our supergravity solutions.  As we have reviewed in \S\ref{sec:intro}, the number of degrees of freedom of this theory scales like $N_c^{\frac{5}{2}}$ at large $N_c$ and hence \eqref{eqn:dualstress} should scale with the same power of $N_c$. One way to see that this is indeed the case is the following.  Consider the string frame metric of our solution in 10D massive type IIA supergravity \cite{Cvetic:1999un}:\footnote{The 10D solution has a curvature singularity that is not hidden behind the horizon. However, for large $N_c$ there is  large a region of the spacetime with $\sin\alpha\gg N_c^{-3/10}$ where both the curvature and the dilaton are small and thus one can trust the supergravity solutions \cite{Brandhuber:1999np,Bergman:2012kr}.} 
\begin{equation}
\dd s^2_{(10)} = \frac{1}{(\sin\alpha)^{\frac{1}{3}}}\big[\ell^2 \dd s_{(6)}^2 + \textstyle{\frac{4\,\ell^2}{9}}(\dd \alpha^2+\cos^2\alpha\,\dd s^2_{S^3})\big]\,.
\end{equation}  
where $\dd s^2_{(6)}$ is our 5D asymptotically AdS solution and $\alpha \in (0,\pi/2]$, so that only the upper hemisphere of a full $S^4$ is covered.  Of the other supergravity fields we will only need the dilaton:\footnote{In field theory, a necessary condition for a fixed point to exist is $N_f\leq 7$ \cite{Seiberg:1996bd}. In supergravity, the solution describing the D-brane system only makes sense if this condition is satisfied \cite{Brandhuber:1999np}.}
\begin{equation}
e^{-2\phi} = \frac{3(8-N_f)^{\frac{3}{2}}\sqrt{N_c}}{2\sqrt 2 \pi}\, (\sin\alpha)^{\frac{5}{3}}\,.
\end{equation}
Ref. \cite{Bergman:2012kr} showed that the relation between the radius of AdS$_6$, $\ell$, and the string length for the class of theories that we are considering is 
\begin{equation}
\frac{\ell^4}{\ell_s^4} = \frac{18\,\pi^2\,N_c}{8-N_f}\,.
\label{eqn:plancklength}
\end{equation}
and the 10D Newton's constant is taken to be $2\kappa_{10}^2=(2\pi)^7g_s^2\ell_s^8$.\footnote{Recall that $2\,\kappa_d^2 = 16\,\pi\,G_d$.} In the decoupling limit the low energy excitations are sensitive to the string frame metric and therefore it seems natural to relate the 10D Newton's constant to the 5D one by dimensional reduction on the $S^4$ in the string frame. By doing so we take into account the fact that the dilaton is non-trivial. This gives the following relation between the 10D Newton constant and the 5D one:
\begin{equation}
\frac{1}{2\,\kappa_6^2} = \frac{1}{2\,\kappa_{10}^2}\,\frac{3(8-N_f)^{\frac{3}{2}}\sqrt{N_c}}{2\sqrt 2 \pi}\,\int \dd\alpha\wedge\textrm{vol}_{S^3}\,\sin\alpha\,\cos^3\alpha\,.
\end{equation}
Combining the result of this integral with \eqref{eqn:plancklength} gives
\begin{equation}
\frac{\ell^4}{16\,\pi\,G_6} = \frac{27}{40\sqrt 2\,\pi^2}\frac{N_c^{\frac{5}{2}}}{\sqrt{8-N_f}}\,,
\end{equation}
which shows that indeed \eqref{eqn:dualstress} scales like $N_c^{\frac{5}{2}}$.

Now we proceed to extract the expectation value of the stress tensor of the dual field theory using holographic renormalisation. Since we are interested in spacetimes with $\mathbb R_t\times$SU$(2)\times$U$(1)$ isometry group, we can consistently truncate the general form of the line element in Fefferman--Graham (FG) coordinates to this particular class of metrics. Solving the Einstein equations in the usual near boundary expansion we find that the stress tensor is given by
\begin{equation}
\langle T_{ij}\rangle\,\dd x^i\dd x^j = \frac{5\,\ell^4}{16\,\pi\,G_6}\,\big[T_5(R)\,\dd t^2 + R_5(R)\,\dd R^2 + B_5(R)(\sigma^3 - \Omega_5(R)\,\dd t)^2 + S_5(R)\,\dd\Omega_{(2)}^2\big]\,,
\label{eqn:stresstensortmp}
\end{equation}
where $T_5,\,R_5,\,B_5,\,\Omega_5,\,S_5$ can be extracted from our numerical solutions. These functions are not all independent: the Einstein equations at 5th order in the near boundary expansion (in FG coordinates) impose an algebraic constraint between them which is nothing but the tracelessness condition of the dual stress tensor.\footnote{Recall that in odd number of boundary dimensions, the stress tensor is always traceless regardless of the boundary metric being Ricci flat or not because there is no gravitational conformal anomaly.} At 6th order one finds a differential constraint which implies that \eqref{eqn:stresstensortmp} is covariantly conserved. Therefore, we find that the stress tensor is fully specified in terms of three independent functions. However, in our construction we shall independently extract the five functions from our numerical data, then use the trace and divergence of $\langle T_{ij} \rangle$ to estimate the associated numerical error. We find that the trace is largest near the horizon, with magnitude of around $10^{-5}$ to $10^{-3}$ (the fast rotating cases giving smaller values). This quickly decays to well below $10^{-9}$ in all cases as we move away from the horizon.

In order to obtain the components of the stress tensor in terms of our numerical solutions we have to change from the working coordinates $(x,r)$ used in our ansatz \eqref{eqn:ansatz} into the Fefferman--Graham ones, $(z,R)$ that we used to get \eqref{eqn:stresstensortmp}. We proceed as in \cite{Figueras:2012rb} and determine the change of coordinates $z=z(x,r),\,R=R(x,r)$ in a near boundary expansion, requiring that $\xi^a=0$ at each order in $\left(1-x^2\right)$.  Note that in our setup $\xi^a$ has only two non-vanishing components so imposing $\xi^a=0$ order by order determines the coordinate change completely.

The dual stress tensor $\langle T_{ij}\rangle$ is symmetric in its two indices but because the boundary metric \eqref{eqn:MP5tmp} is not positive definite, the linear map $\langle T^i_{\phantom i j }\rangle$ from vectors to vectors need not be diagonalisable. However, in our case we find that it is, and for all values of the angular momentum parameter $a$ of the boundary 5D Myers--Perry black hole that we have considered there is one timelike eigenvector and four spacelike eigenvectors. We can identify the corresponding eigenvalues with the energy density and the pressures in the following way. We write the stress tensor as
\begin{equation}
\langle T^i_{\phantom i j}\rangle = \rho(R)\,t^i\otimes t_j + p_1(R)\big[(s^1)^i\otimes (s^1)_j + (s^2)^i\otimes (s^2)_j \big] + p_2(R)\,(s^3)^i\otimes (s^3)_j + p_3(R)\,R^i\otimes R_j 
\label{eqn:stresstensor}
\end{equation}
where $t^i$ is the unique timelike eigenvector normalised so that $t^it_i=-1$, and $(s^\flat)^i$ and $R^i$ are the orthonormal spacelike vectors in the 5D Myers--Perry background\footnote{We choose $(s^\flat)_i \sim (\sigma^\flat)_i$ with the obvious proportionality factors and $R_i = \sqrt{\frac{R^2 (R^2 + a^2)}{(R^2 + a^2)^2 - \mu R^2}}\,(dR)_i$.} written in Boyer--Lindquist-type  coordinates, \eqref{eqn:MP5tmp}.  In this way we identify $\rho(R)$ as the energy density of the plasma seen by a local observer with velocity $t^i$, and the $p_i(R)$ as the corresponding pressures.\footnote{Recall that if $k^i$ is a unit timelike and future directed vector field, then $-T^i_{\phantom i j}\,k^j$ represents the energy density seen by an observer with velocity given by $k^a$. So $-\rho(R)$ is the eigenvalue of the matrix $T^i_{\phantom i j}$.}

\begin{figure}[t]
\begin{center}
\includegraphics[scale=0.45]{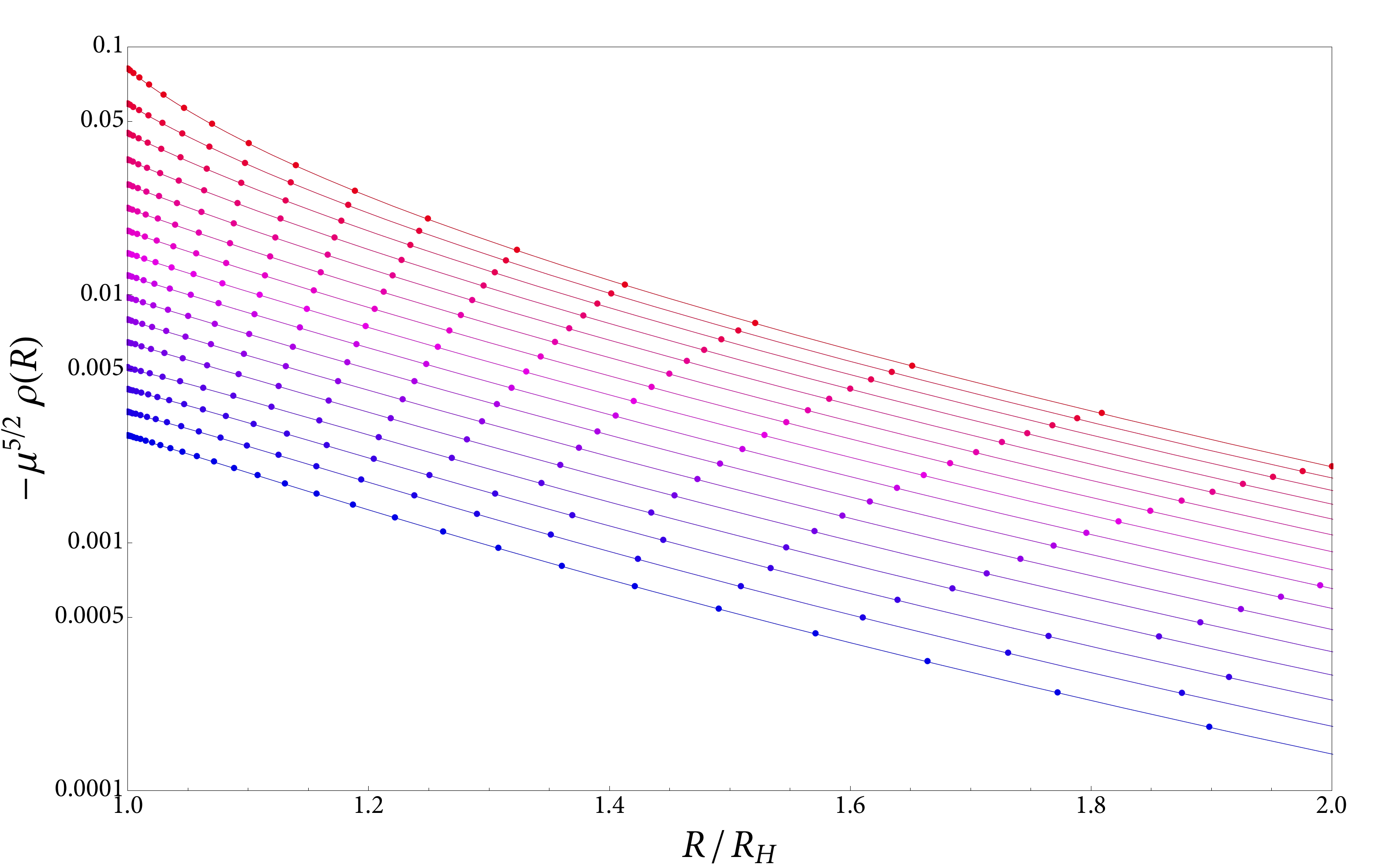}
\includegraphics[scale=0.275]{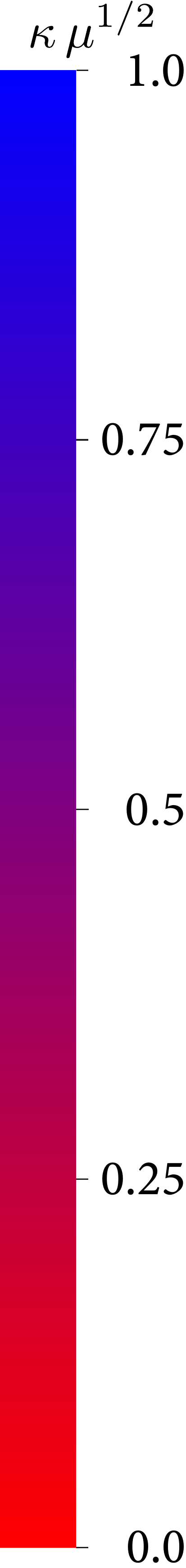}
\end{center}
\caption{Plot of (minus) the energy density for the different values of the rotation parameter versus the Boyer--Lindquist radial coordinate $R$.  We have used a logarithmic scale on the $y$ axis in order to enhance the near horizon region. As this plot shows, the energy density is negative everywhere and it becomes more negative as the rotation of the black hole increases.}
\label{fig:timeeigen}
\end{figure}

In Fig. \ref{fig:timeeigen} we plot the energy density for different values of the rotation parameter of the background.  First note that the energy density is \textit{negative} everywhere in our domain for all $a\geq 0$. Moreover, the energy density is a monotonically increasing function of the radial coordinate (\textit{i.e.,} it becomes less negative for large $R$) and it is largest (in absolute value) at the horizon. Further, as the angular momentum increases (moving up in this figure) the energy density becomes more negative at the horizon, but it is always finite. By continuity the energy density should also be finite at extremality. 
Next we consider the pressures (see Fig. \ref{fig:pressures} in Appendix \ref{sec:pressures} for the plots). As in \cite{Figueras:2012rb} we find that the radial pressure is positive whilst the pressures along the angular directions are negative, and therefore there is an anisotropy in the system. Note that this anisotropy is already present in the static limit. Adding  rotation to the boundary black hole breaks the full SO$(4)$ spherical symmetry of the static case down to SU$(2)\times$U$(1)$, which results in some further anisotropy between the fibre and  the sphere directions. This simply reflects the squashing of the $S^3$ of the boundary black hole. As angular momentum of the boundary black hole increases, the horizon $S^3$ becomes more squashed which gives rise to a more negative  pressure along the fibre direction than  along the base $S^2$ directions.

\begin{figure}[t]
\begin{center}
\includegraphics[scale=0.5]{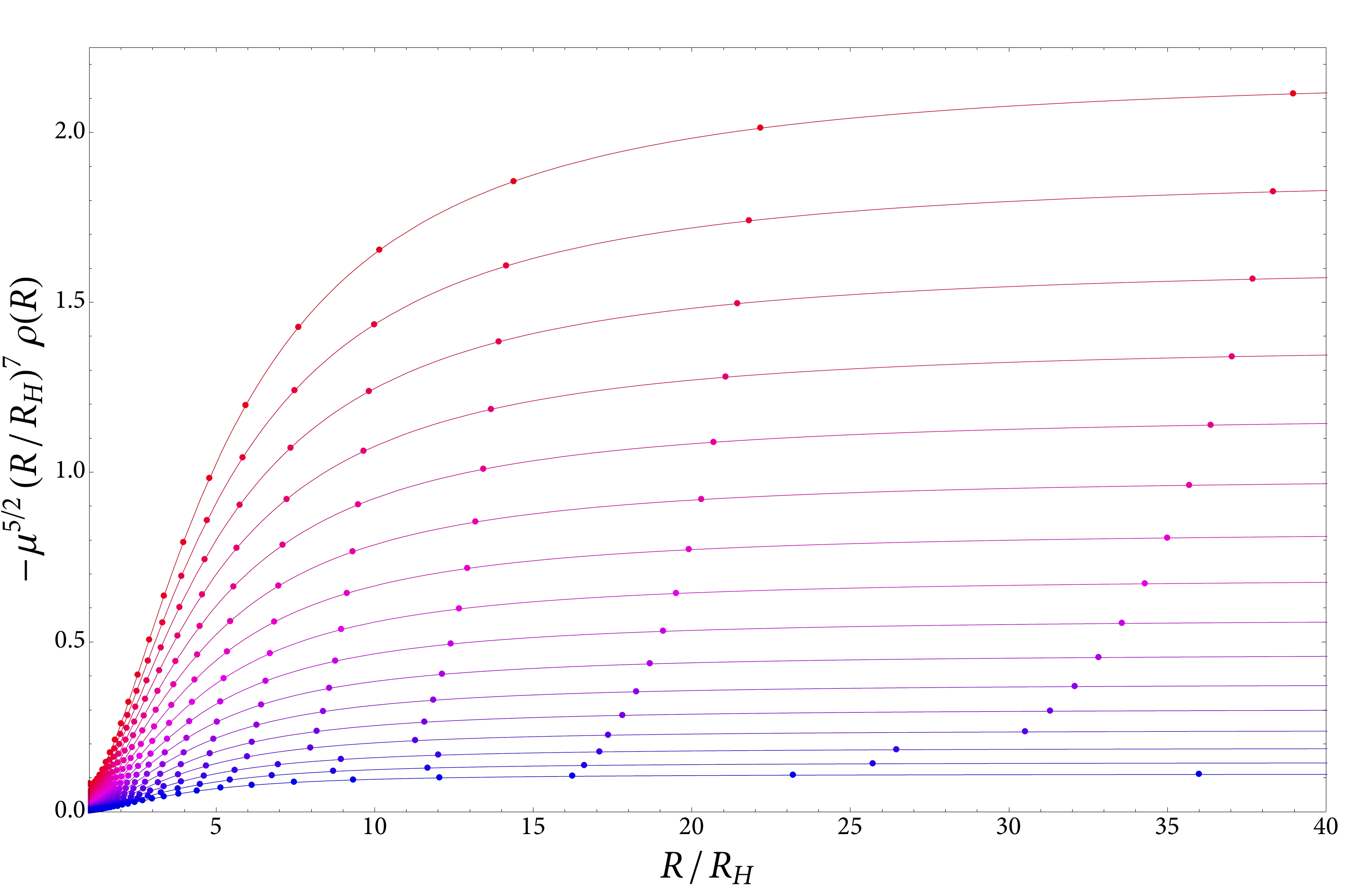}
\end{center}
\caption{Plot of the rescaled energy density for the different values of the rotation parameter. This plot shows that the energy density decays like $1/R^7$ in the asymptotic $R\to\infty$ region.  The other components of the stress tensor have a similar behaviour near infinity. The colour code is the same as in Fig.\ref{fig:timeeigen}.}
\label{fig:eigenlargeR}
\end{figure}

In Fig. \ref{fig:eigenlargeR} we study the behaviour of the energy density as a function of the radial coordinate $R$ in the $R\to\infty$ region. As one can see from this figure,  the energy density decays like $1/R^7$ near infinity and the other components of the stress tensor exhibit the same asymptotic behaviour (see Fig. \ref{fig:rescaledpressures} in Appendix \ref{sec:pressures}).  Not surprisingly, this can be understood as in the 4D case \cite{Tanahashi:2011xx}. The energy density stored in the field around a point particle in 5-spacetime dimensions behaves like $\propto N_c^{\frac{5}{2}}\,R_g/R^7$, where $R_g$ is the gravitational radius of the point particle.  The pressures are comparable in magnitude to the energy density and they also decay like $1/R^7$.  
In Appendix \ref{sec:linear} we recover this result by adapting the linearised calculation of \cite{Garriga:1999yh} to the present situation, where the bulk spacetime is 6-dimensional.

\begin{figure}[t]
\begin{center}
\includegraphics[scale=0.5]{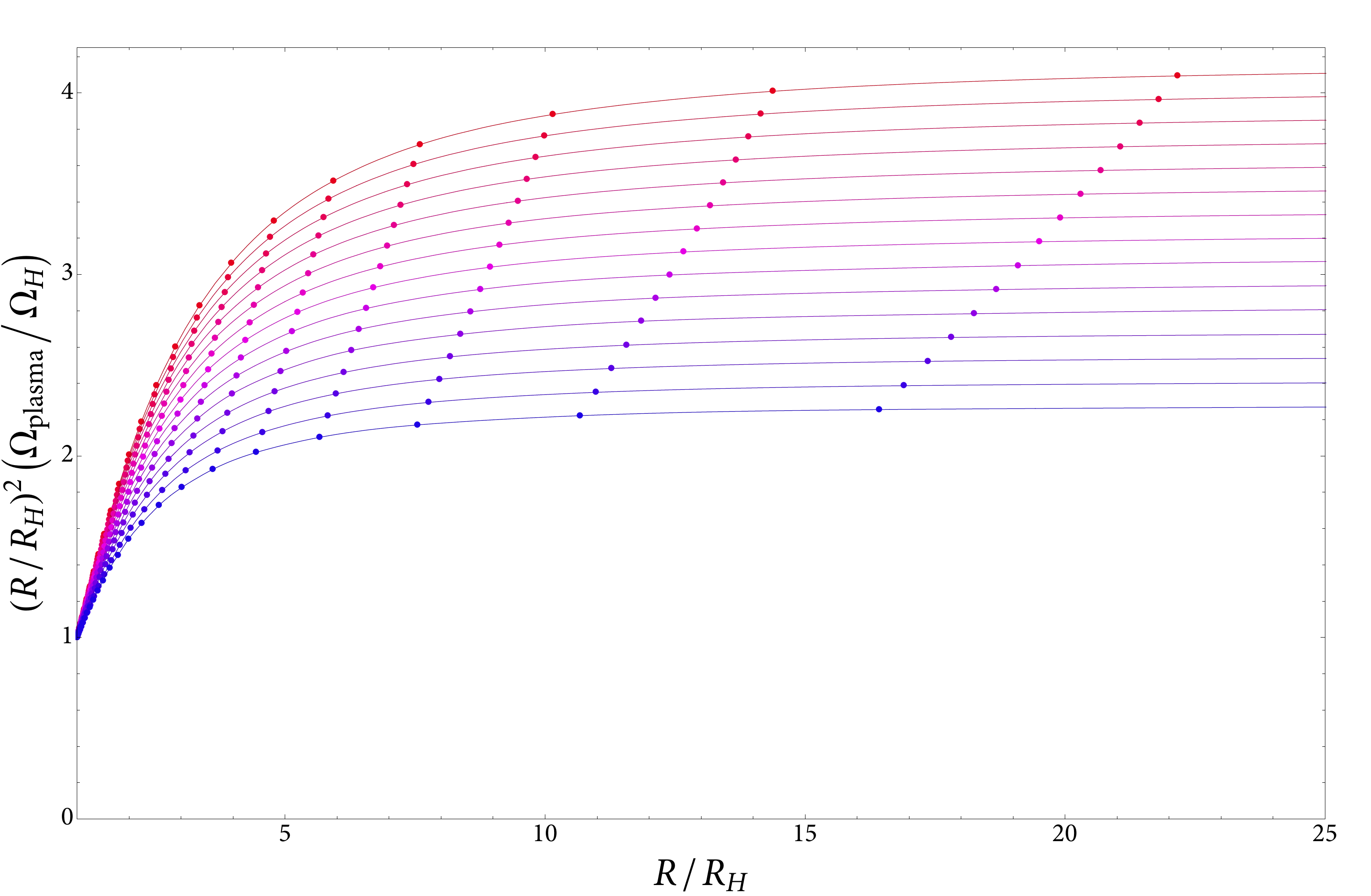}
\end{center}
\caption{Rescaled angular velocity of the plasma as  function of the radial coordinate $R$ for the  different values of the rotation parameter.  Near infinity the angular velocity of the plasma decays as $\sim 1/R^2$ and at the horizon it reduces to the angular velocity of the black hole.}
\label{fig:angvel}
\end{figure}

Finally we consider the rotation of the dual plasma. To do so, we write the unique timelike eigenvector of the stress energy tensor of the dual CFT as
\begin{equation}
T = \frac{\partial}{\partial t} + \Omega(R)\,\frac{\partial}{\partial \psi}\,.
\label{eqn:timelikeeigen}
\end{equation}
This allows us to identify  $\Omega(R)$ as the angular velocity of the plasma  with respect to a static observer at infinity in the boundary directions. In Fig. \ref{fig:angvel} we plot this quantity for different values of the angular momentum parameter $a$.  As one might have expected, $\Omega(R) \to \Omega_H$ as $R\to R_H$. That is, at the horizon the plasma is co-rotating with the black hole. This is an example of frame dragging even though at the boundary gravity is non-dynamical. This had to be the case since the boundary black hole is rotating rigidly with respect to infinity; as there is no flux of CFT plasma through the horizon then it must be that the plasma is co-rotating with the black hole.  In the asymptotic region, $R\to \infty$, we find that the angular velocity of the plasma decays like $1/R^2$, so that at infinity the plasma is at rest and the timelike eigenvector coincides with the asymptotic timelike Killing vector.

We now discuss the physical interpretation of our results.  From Fig. \ref{fig:timeeigen}  we see that the energy density of the plasma is negative. Because the dual stress tensor is traceless, the weak and the strong energy conditions are equivalent in this case and therefore \eqref{eqn:stresstensor} violates the usual energy conditions considered in classical general relativity. The same is true for the stress energy tensor of $\mathcal N=4$ SYM in the background of 4D Schwarzschild in the Unruh vacuum computed in \cite{Figueras:2011va}. However, the energy density in the Hartle--Hawking state computed in \cite{Santos:2012he} turns out to be positive. One may be concerned that the fact that we are seeing a negative energy density may be a sign of a pathology and/or an instability of these solutions. However, this need not be the case since our stress tensor corresponds to that of a \textit{quantum} field theory and as such it need not satisfy any of the classical energy conditions. 

 Ref. \cite{Figueras:2011va} (see also \cite{Tanahashi:2011xx}) provided a heuristic physical interpretation of the stress tensor \eqref{eqn:stresstensor} as a `halo' of plasma in equilibrium with the boundary black hole. According to this picture,  the strong attractive self-interactions of the CFT are balanced by thermal radiation pressure from the black hole.  There is another instance in which putting a QFT in a non-trivial background results in a stress tensor with some properties that are qualitatively similar to the ones we find.  This is the case of the celebrated Casimir effect \cite{Casimir:1948dh}, in which the vacuum energy of the electromagnetic field between two perfectly conducting plates is negative, giving rise to an attractive force between the two plates. Therefore, the fact that the energy density in \eqref{eqn:stresstensor} is negative may simply reflect the attractive nature of the self-interactions of the CFT and the non-vanishing vev of the stress tensor may be due to the Casimir effect.\footnote{Ref. \cite{Emparan:2002px} noted a large contribution of the Casimir effect in the stress tensor of a conformally coupled weakly interacting scalar field in the background of the BTZ black hole. }  

We may now try to better understand \eqref{eqn:stresstensor} in terms of the Casimir effect or equivalently vacuum polarisation. To do so we may concentrate on the $t$-$r$ part of the geometry since this gives rise to the non-trivial causal structure of the spacetime: the sphere directions are mere spectators in both this paper and in \cite{Figueras:2011va} because of symmetry assumptions. Therefore, we may try to understand the behaviour of the $t$-$t$ and $r$-$r$ components of the stress tensor by considering a simple 2D model.  For concreteness we will consider $\langle T^i_{\phantom i j}\rangle$ for a generic quantum field theory in the 2D space
\begin{equation}
\dd s^2 = - \left(1-\frac{2\,M}{r}\right)\,\dd t^2 + \frac{\dd r^2}{1-\frac{2\,M}{r}}\,,
\label{eqn:2dSchw}
\end{equation}
where we have taken the $t$-$r$ part of the 4D Schwarzschild solution but our considerations should apply without change to the $t$-$r$ part of our geometry.  The advantage of working in 2D is that the stress tensor is completely determined by symmetries and conservation in terms of purely geometrical data \cite{Davies27061977,Christensen:1977jc}. Indeed, assuming that the stress tensor is  time-symmetric, one finds that the conservation equation reduces to
\begin{equation}
\partial_r\langle T^r_{\phantom r r}\rangle  = \frac{M}{r^2}\frac{1}{1-\frac{2\,M}{r}}\big(\langle T^t_{\phantom t t}\rangle - \langle T^r_{\phantom r r} \rangle\big) \quad \Rightarrow \quad \partial_r\left[\left(1-\frac{2\,M}{r}\right)\langle T^r_{\phantom r r}\rangle\right] = \frac{M}{r^2}\,\langle T^i_{\phantom i i}\rangle\,.
\end{equation}
This equation can be readily integrated to get
\begin{equation}
\langle T^i_{\phantom i j}\rangle = \textrm{diag}\left(-\frac{H_2(r)}{1-\frac{2\,M}{r}} + \langle T^i_{\phantom i i}\rangle,\,\frac{H_2(r)}{1-\frac{2\,M}{r}} \right)\,,
\label{eqn:stress2dtmp}
\end{equation}
where the first entry denotes the $t$-$t$ component of the stress tensor and
\begin{equation}
H_2(r) = M\,\int_{2M}^r\,\dd r' \frac{\langle T^i_{\phantom i i}\rangle(r')}{r'^2}\,.
\label{eqn:h2}
\end{equation}
Therefore, we see that the full stress tensor is completely determined by its trace. The latter is fixed by the conformal anomaly, which in 2D and for \eqref{eqn:2dSchw} gives
\begin{equation}
\langle T^i_{\phantom i i}\rangle = \alpha \, R = \frac{4\,\alpha\,M}{r^3}\,,
\end{equation}
where $\alpha$ is a constant that depends on the spin of the field being considered. For a free massless scalar field one has $\alpha = \frac{1}{24\,\pi}$ \cite{Davies27061977,Christensen:1977jc}. Using this in \eqref{eqn:h2} allows us to completely determine the expectation value of the full  stress tensor in this geometry:
\begin{equation}
\langle T^i_{\phantom i j}\rangle = \alpha\,\textrm{diag}\left(\frac{4\,M}{r^3} - \frac{1}{16\,M^2}{\textstyle\big(1+\frac{2M}{r}\big)\big(1+\frac{4M^2}{r^2}\big)},\,\frac{1}{16\,M^2}{\textstyle\big(1+\frac{2M}{r}\big)\big(1+\frac{4M^2}{r^2}\big)}\right)\,.
\label{eqn:stress2d}
\end{equation}

In the near horizon region \eqref{eqn:stress2d} has the same qualitative features as our 5D stress tensor or that of \cite{Figueras:2011va}. In particular, it is static and regular on both the future and the past event horizons, and the energy density near the horizon is \textit{negative}. Clearly, the origin of \eqref{eqn:stress2d} is due to the spacetime curvature near the horizon and hence one may conclude that it is a vacuum polarisation effect.  Of course, for large $r$ \eqref{eqn:stress2d} reduces to the stress tensor of radiation at the Hawking temperature set by the horizon in \eqref{eqn:2dSchw}, and clearly this is not what we see. Therefore, this 2D model captures only the physics in the near horizon region but not in the asymptotic region near infinity.  

Now consider adding the extra dimensions  to take into account the full 4D Schwarzschild geometry. In this case, we shall assume that the stress tensor is static, spherically symmetric and traceless because we are interested in considering CFTs in this background. With these assumptions, the conservation of the stress tensor fully determines it up to one function \cite{Christensen:1977jc,Figueras:2011va}:
\begin{equation}
\begin{aligned}
&\langle T^i_{\phantom i j}\rangle = \textrm{diag}\left(-\frac{H_4(r)}{r^2(1-\frac{2M}{r})}-2\,\Theta(r),\,\frac{H_4(r)}{r^2(1-\frac{2M}{r})},\,\Theta(r),\,\Theta(r)\right)\,,\\
&H_4(r)=2\int_{2M}^r\dd r'(r'-3\,M)\Theta(r')\,,
\end{aligned}
\label{eqn:stress4d}
\end{equation}
where the first two entries in \eqref{eqn:stress4d} correspond to the $t$-$t$ and $r$-$r$ components and $\langle T^\theta_{\phantom \theta\theta}\rangle = \langle T^\phi_{\phantom \phi\phi}\rangle = \Theta(r) $ by spherical symmetry.  Adding the extra dimensions introduces new pressures along the angular directions and the extra powers of $r$ in the denominators in \eqref{eqn:stress4d}  can alter the behaviour of the components of the stress tensor near infinity.

The function $\Theta(r)$ in \eqref{eqn:stress4d}  determines the full stress tensor and its actual form depends on details of the CFT under consideration.  However, one can put some constraints on it on general grounds, without assuming strong coupling. First, note that in order for \eqref{eqn:stress4d} be finite at the horizon $(r=2\,M)$, we only need to require that $\Theta |_{r=2M}$ be finite. In addition, we have seen above that the 2D model captures the vacuum polarisation effect near the horizon and therefore we need to require that $\langle T^t_{\phantom tt}\rangle >0$ for $r\approx 2\,M$. This implies that $\Theta<0$ in this region and hence we are led to conclude that the polarisation of the vacuum near the horizon will inevitably make the pressures along the angular directions negative. Also note that from \eqref{eqn:stress4d} one concludes that at the horizon $r=2\,M$ the following relations must hold: $\langle T^t_{\phantom tt}\rangle|_{r=2M}= \langle T^r_{\phantom rr}\rangle|_{r=2M} = -\Theta|_{r=2M}$. A close examination of the data of \cite{Figueras:2011va} shows that this is true for $\mathcal N=4$ SYM in the background of Schwarzschild, the deviations being smaller than $1\%$. At infinity we want $\langle T^i_{\phantom i j}\rangle \to 0$ and for this we need $\Theta \sim 1/r^{3}$ or faster. Then \eqref{eqn:stress4d} implies that all the components of the stress tensor have the same behaviour near infinity. Note that $\Theta$ has to decay sufficiently fast if we want that $\langle T^t_{\phantom tt}\rangle$ and $\langle T^r_{\phantom rr}\rangle$ do not change sign at a sufficiently large $r$. One can infer  the fall off of $\Theta$ from the linearised calculation of \cite{Garriga:1999yh}  and one finds that the pressures along the sphere directions are negative and the fall off is   $ \sim 1/r^5$.

Finally we note that even though we find negative energy densities,   quantum energy inequalities (see \cite{Fewster:2012yh} for a review with references) can be  used to bound them. Quantum energy inequalities are the  remnants of the classical energy conditions satisfied by quantum field theories and they quantify the extent to which a quantum field can violate these classical energy conditions. Even though most of the results available in the literature apply to Minkowski spaces, on physical grounds one expects that similar results hold on curved spacetimes over regions which are small compared the curvature length scales. In fact,   recent progress  \cite{Fewster:2006kt} shows that this intuition is indeed correct.     For instance, consider a globally hyperbolic spacetime $\mathcal M$ with a timelike Killing field $t^a$ admitting a foliation into constant time surfaces such that $\mathcal M\simeq \mathbb R_t\times \Sigma$. For any region $\Sigma_0\subset \Sigma$ such that the spacetime metric takes the Minkowski form, the energy density of a free scalar field satisfies \cite{Fewster:2006kt}\footnote{Interactions can also be included and they do not change in essence the quantum energy inequalities, \cite{Fewster:2012yh}.}
\begin{equation}
\langle T_{ab}\,n^a\,n^b\rangle_\omega\geq -\frac{C_4}{(2\,r(x))^4}\,,
\end{equation}
for any Hadamard state $\omega$, and where $n^a$ is the unit vector along $t^a$. Here $r(x)$ is the radius of the largest Euclidean 3-ball which can be isometrically embedded in $\Sigma_0$ and $C_4=3.169858...$, is a constant. For Dirac or vector fields one can obtain similar type of inequalities. Our results, obtained using AdS/CFT and valid at strong coupling, show that in this regime the energy density also remains bounded from below.

\section{Conclusions}  
\label{sec:conclusions}

In this paper we have constructed the gravitational duals of 5D CFTs in rotating black hole backgrounds. For simplicity we have considered the case in which the boundary black hole is the 5D Myers--Perry solution with equal angular momenta because the metric is of co-homogeneity one, and hence the full spacetime is of co-homogeneity two. Moreover, in the region far from the horizon the spacetime reduces to the Poincar\'e horizon of AdS and hence the CFT should be in the Unruh state. Whilst our solutions should capture the general qualitative physics of the Unruh vacuum in rotating black hole backgrounds (at large $N_c$ and strong coupling) it would still be interesting to consider the gravitational dual of $\mathcal N=4$ SYM on the background of Kerr. From a computational viewpoint the problem reduces to solving elliptic PDEs in 3 variables and we hope to address it in the future.  The construction of the gravitational duals of CFTs on rotating black hole backgrounds becomes even more interesting if one insists on considering an IR horizon with finite temperature. As \cite{Santos:2012he} pointed out, this requires the horizon of the black hole be non-Killing and recently methods for constructing this kind of solutions have been developed \cite{Figueras:2012rb,Fischetti:2012vt}.

We have also noted that a generic feature of the Unruh vacuum for strongly coupled CFTs in black hole backgrounds is that the energy density is negative everywhere in the domain of outer communications. We have argued that this can be understood in terms of the polarisation of the vacuum due to the curvature of the background. Further, our stress tensor \eqref{eqn:stresstensor} is bounded from below, in accordance generic local quantum energy inequalities \cite{Fewster:2012yh}.  The fact that the energy density is negative need not signal that our solutions are unstable but we have not performed a detailed study of their stability  under gravitational perturbations. It would be very interesting to carry out such a study.

It is interesting to compare the results of this paper (and \cite{Figueras:2011va}) with the black funnels \cite{Santos:2012he}. The latter represent the gravitational dual of the Hartle--Hawking state for $\mathcal N=4$ SYM on the background of Schwarzschild. In the black funnels case, the energy density is positive \textit{everywhere} and not only at infinity (see Fig. 4 in  \cite{Santos:2012he}), where the stress tensor reduces to that of pure radiation at the Hawking temperature set by the IR horizon. Therefore, their stress tensor does not seem to have a component corresponding to the vacuum polarisation. This should be contrasted with the free field theory results \cite{Howard:1984qp}, which indicate that near the horizon it is the vacuum polarisation that dominates the stress tensor, as in this paper. However,  \cite{Hubeny:2009ru} conjectured that there should exist another family of solutions which are dual to the Hartle--Hawking state, namely the black droplets. The latter   should arise as a continuous deformation of the solutions of \cite{Figueras:2011va}, where the continuous parameter is the temperature of the IR horizon. Because in \cite{Figueras:2011va} (and of course in our present solutions as well) the non-vanishing stress tensor is due to vacuum polarisation effects, the stress tensor of the black droplets should be dominated by vacuum polarisation effects near the horizon, at least when the temperature of the IR black hole is sufficiently small compared to the size of the boundary black hole. In the asymptotic region, the stress tensor should reduce to pure radiation, as in the black funnels case. Therefore, for the black droplets the $\langle T^t_{\phantom t t}\rangle$ component of the stress tensor should change sign, being positive near that horizon and negative in the near infinity region.  It would be very interesting to construct these solutions.

\subsection*{Acknowledgements}
We are grateful to Sungjay Lee, Nick Dorey, Diego Rodriguez-Gomez and especially Harvey Reall and Toby Wiseman for discussions.  PF supported by an EPSRC postdoctoral fellowship [EP/H027106/1].  ST is supported by the European Research Council grant no. ERC-2011-StG 279363-HiDGR. The computations presented in this paper were done on COSMOS at the UK National Cosmology Supercomputing Centre in Cambridge.

\appendix

 \section{Further details of the stress tensor}
 \label{sec:pressures} 

 In this appendix we collect more results about the two remaining eigenvalues, namely $p_1(R)$ and $p_2(R)$, which together with the energy density $\rho(R)$ discussed in the main text, fully determine the stress tensor. 
 
\begin{figure}
 \begin{center}
 \includegraphics[scale=0.5]{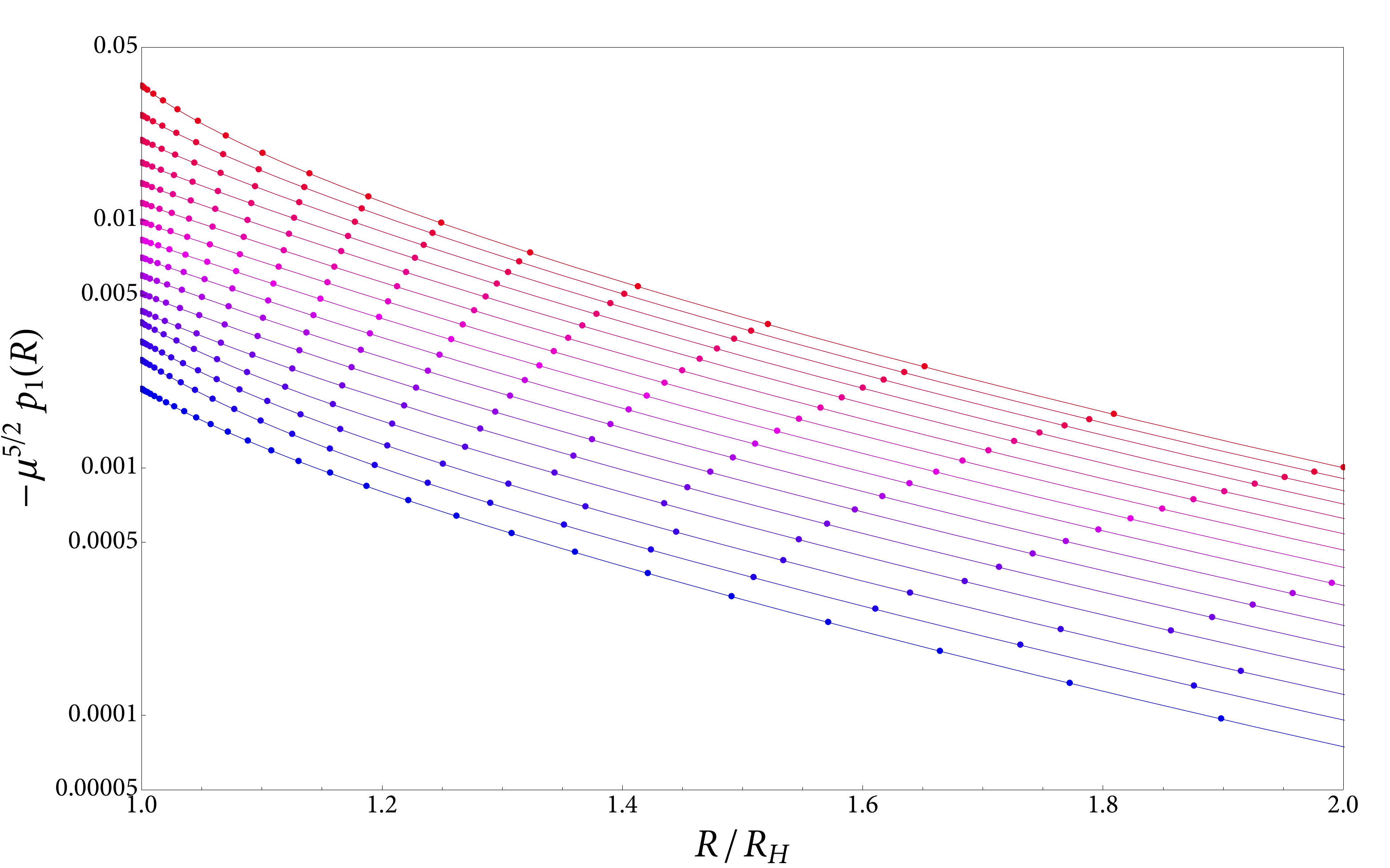}
 \\
 \includegraphics[scale=0.5]{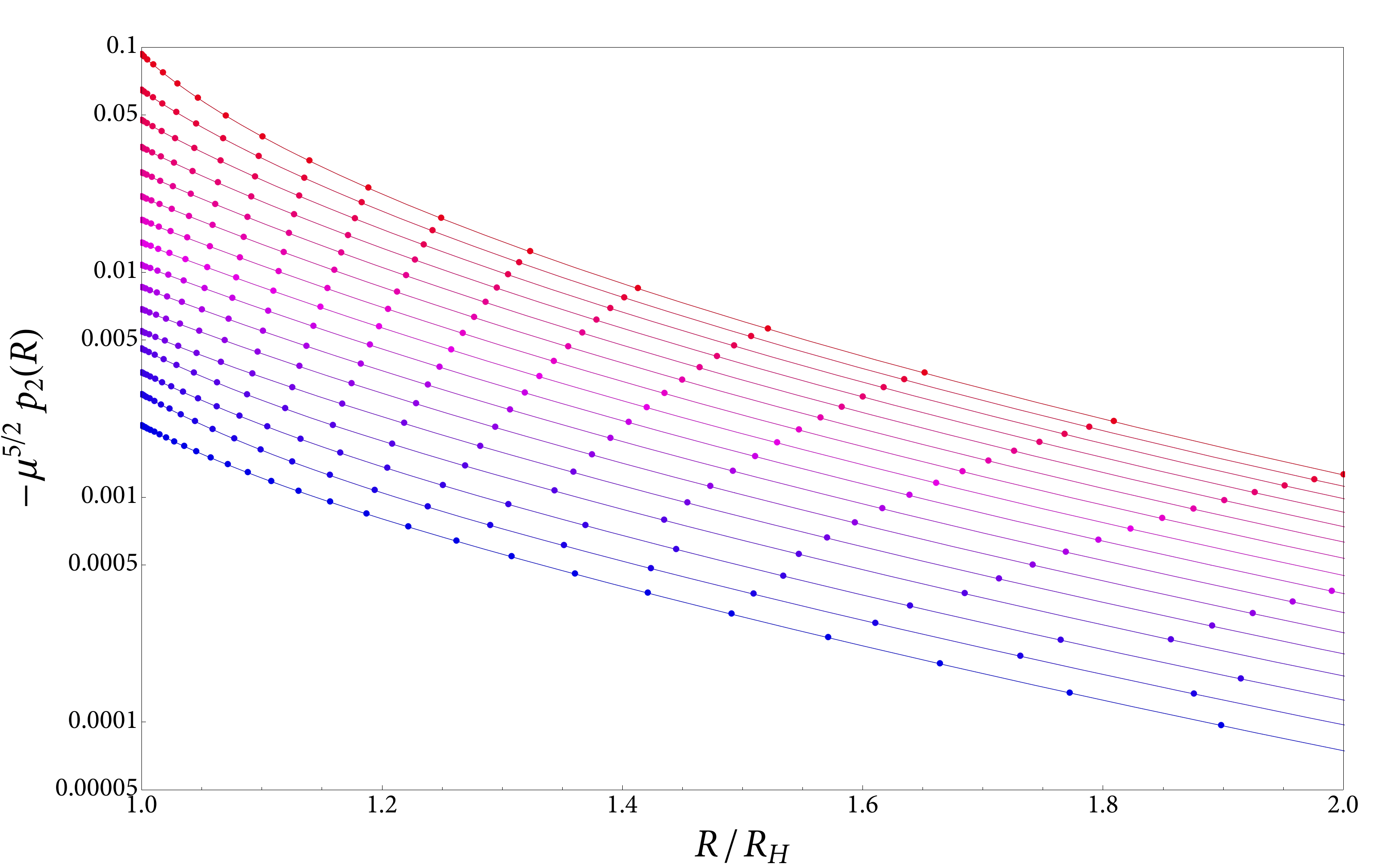} 
 \end{center}
 \caption{Pressures along the sphere directions (\textit{top}) and along the fibre (\textit{bottom}) in the near horizon region. Notice the log scale of the $y$ axis. The pressures are finite and negative everywhere in the domain. }
 \label{fig:pressures}
 \end{figure}

\begin{figure}
 \begin{center}
 \includegraphics[scale=0.5]{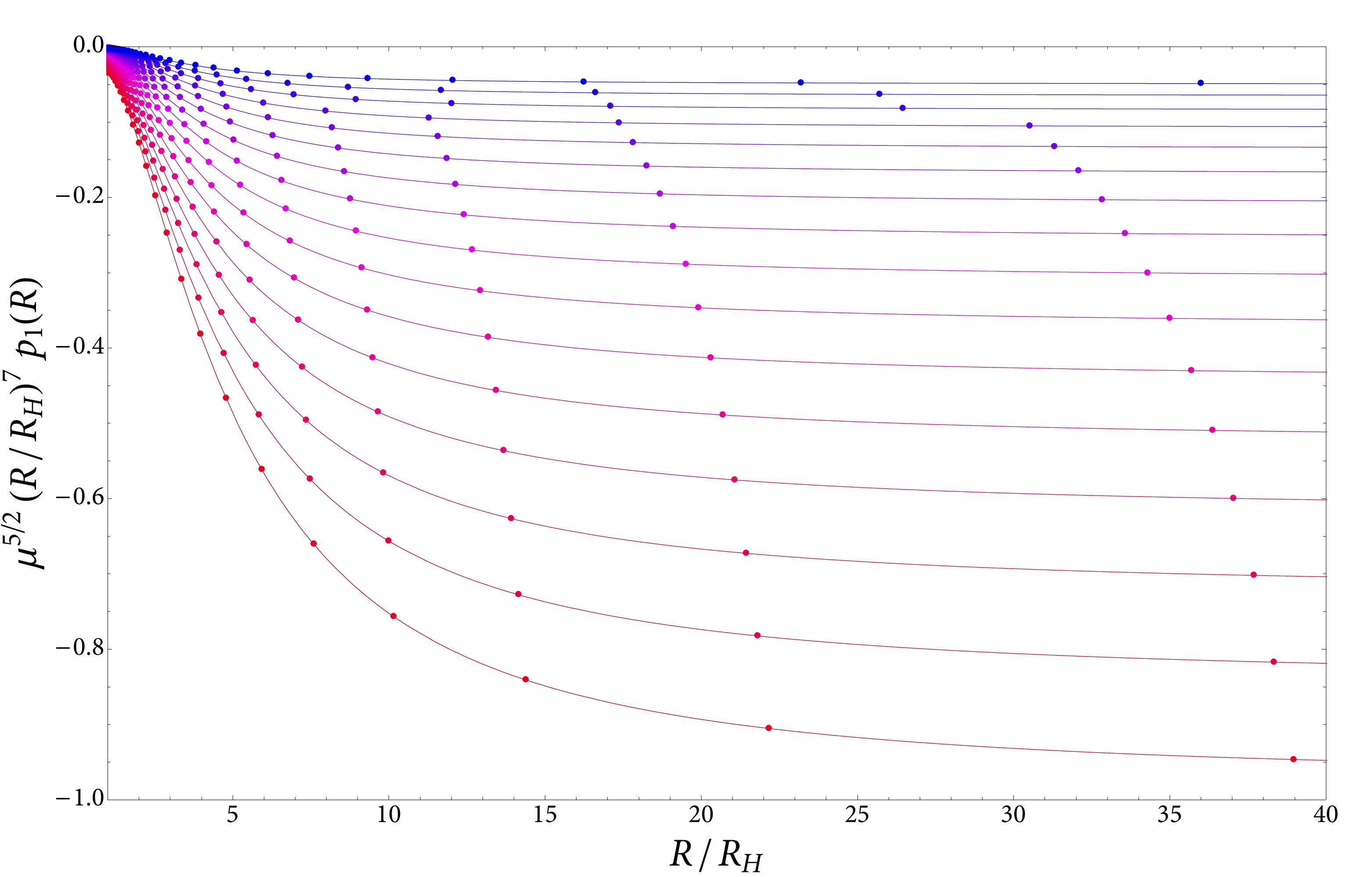}
 \\
 \includegraphics[scale=0.5]{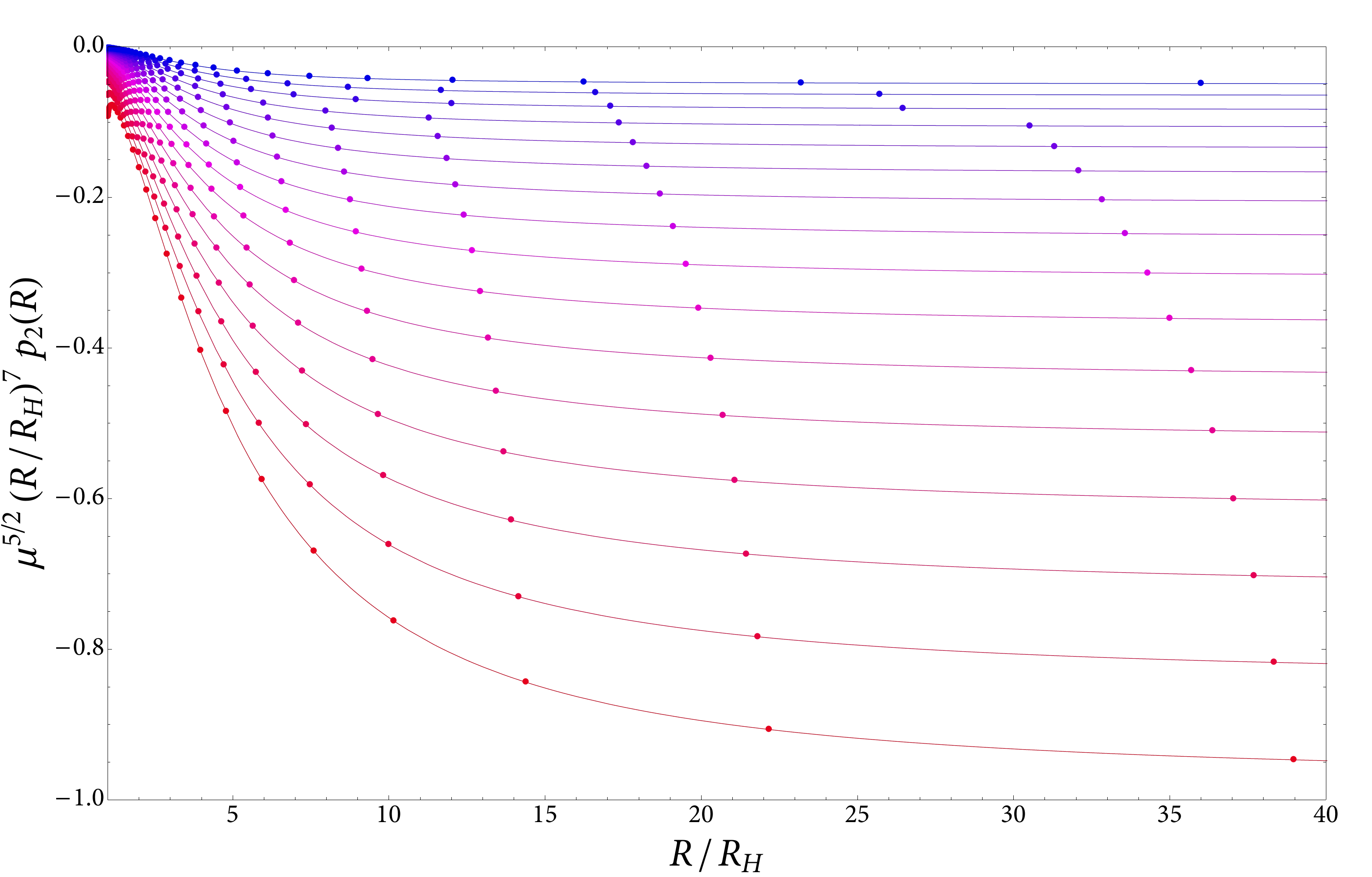} 
 \end{center}
 \caption{Rescaled pressures along the sphere directions (\textit{top}) and along the fibre (\textit{bottom}). Both pressures decay as $1/R^7$ for $R\to\infty$.}
 \label{fig:rescaledpressures}
 \end{figure}

 In Fig. \ref{fig:pressures} we plot the absolute value of the two pressures, although both of them are negative everywhere in our domain (see Fig. \ref{fig:rescaledpressures}). In these plots  we have used a log scale in the $y$ axis so as to zoom into the region near the horizon. As shown in this figure, both pressures are finite at the horizon. It is interesting to note that the pressure along fibre direction is always larger (in absolute value) than the pressure along the $S^2$. In Fig. \ref{fig:rescaledpressures} we plot the rescaled pressures. As this plot shows, both pressures decay as $\sim 1/R^7$ at large distances, as advertised in the main text.

\section{Convergence tests}
\label{sec:convergence}  

When a numerical scheme is employed to solve a continuous problem, it is essential to verify that that the solution so obtained converges appropriately as the continuum limit is approached. For spectral methods, the error should decay exponentially as the number of grid points is increased, provided that the function being approximated is smooth. In this appendix, we present evidence to show that our numerical construction does indeed exhibit this behaviour.

\begin{figure}
 \begin{center}
 \includegraphics[scale=0.5]{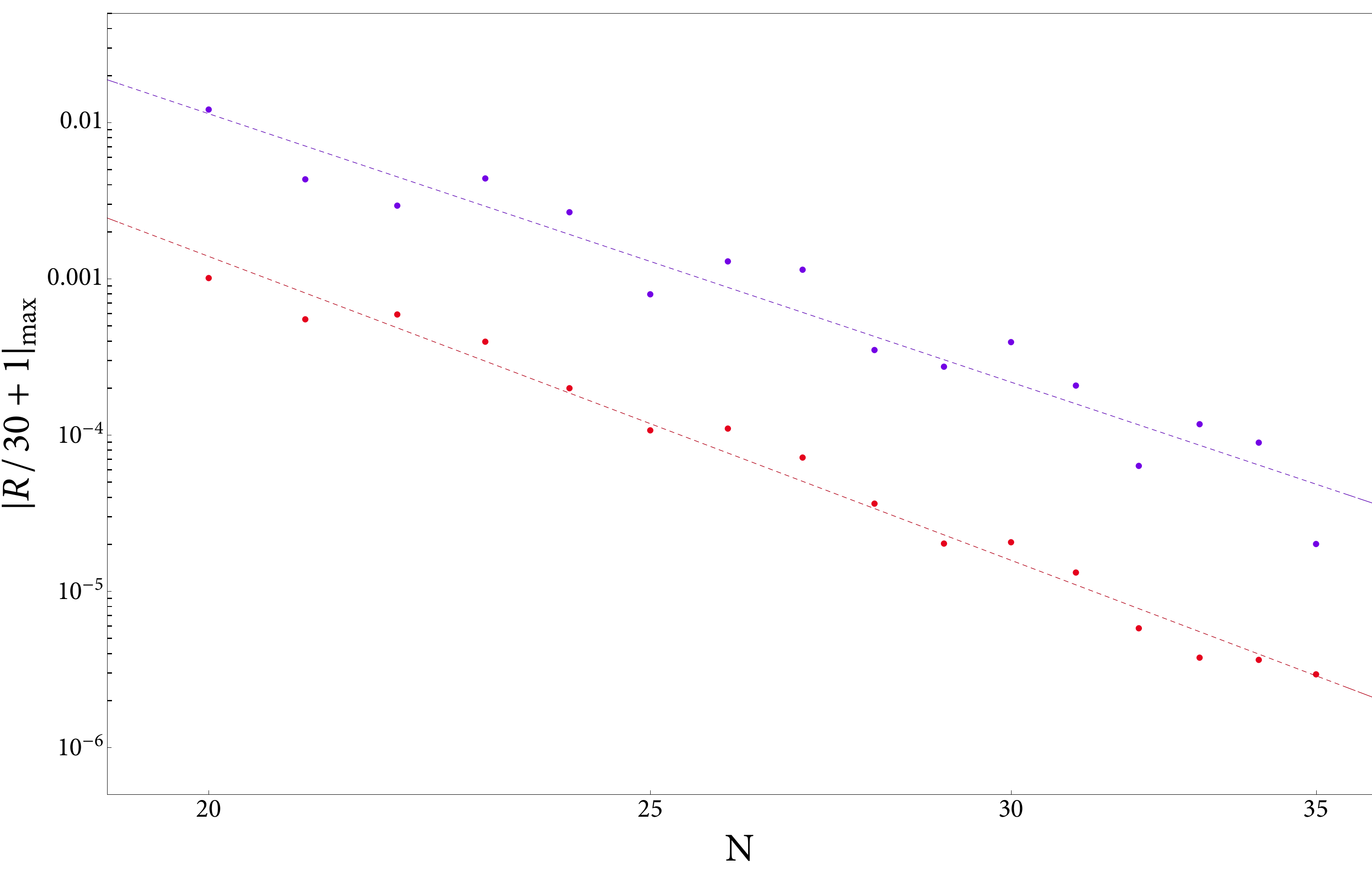}
 \\
 \includegraphics[scale=0.5]{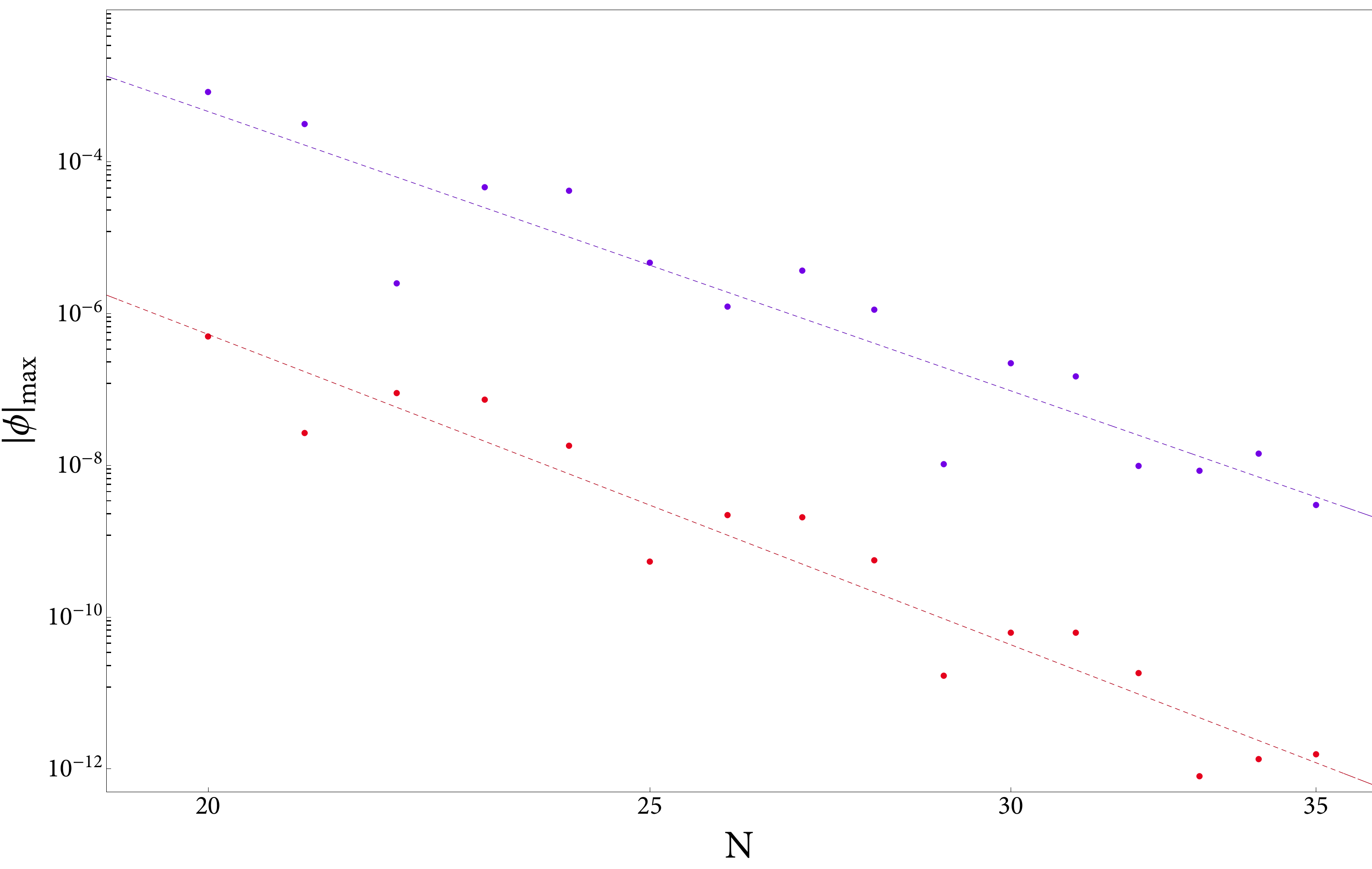} 
 \end{center}
 \caption{Logarithmic plot of $\left|\frac{R}{30}+1\right|_{\textrm{max}}$ and $\left|\phi\right|_{\textrm{max}}$ as a function of the grid size. The red and purple curves correspond to $\kappa\, \mu^{\frac{1}{2}} = 0.0625$ and $0.75$ respectively. In all cases the error decays exponentially, but the more highly-spinning case has a larger decay coefficient.}
 \label{fig:convergence}
 \end{figure}

We seek a metric which satisfies the condition $R_{ab}  = -5\,g_{ab}$, where we have set the AdS radius to one,  $\ell=1$. By taking the trace of this equation, we see that the scalar curvature should approach $R = -30$ in the continuum limit. Therefore, we use the maximum of the quantity $\left|\frac{R}{30} + 1\right|$ as a measure of error in our solution. In Fig. \ref{fig:convergence} we plot this quantity for the metric which is the fixed point of Ricci flow over a range of grid sizes. We performed this convergence test for two parameter values $\kappa \,\mu^{\frac{1}{2}} = 0.0625 , 0.75$. In both cases, we can see that the error follows a general exponentially decaying trend, even though the decay is not monotonic in the grid size.

Furthermore, our \emph{a posteriori} gauge-fixing scheme allows for the possibility that the flow converges towards a Ricci soliton. To check that our solution is indeed Einstein, we also track the decay of the squared-norm of the DeTurck vector: $\phi = \xi^a \xi_a$. Again, we expect that this quantity should become zero in the continuum limit. Fig. \ref{fig:convergence} shows that $|\phi|$ also decays exponentially as we increase the number of grid points.
 
\begin{figure}
 \begin{center}
 \includegraphics[scale=0.5]{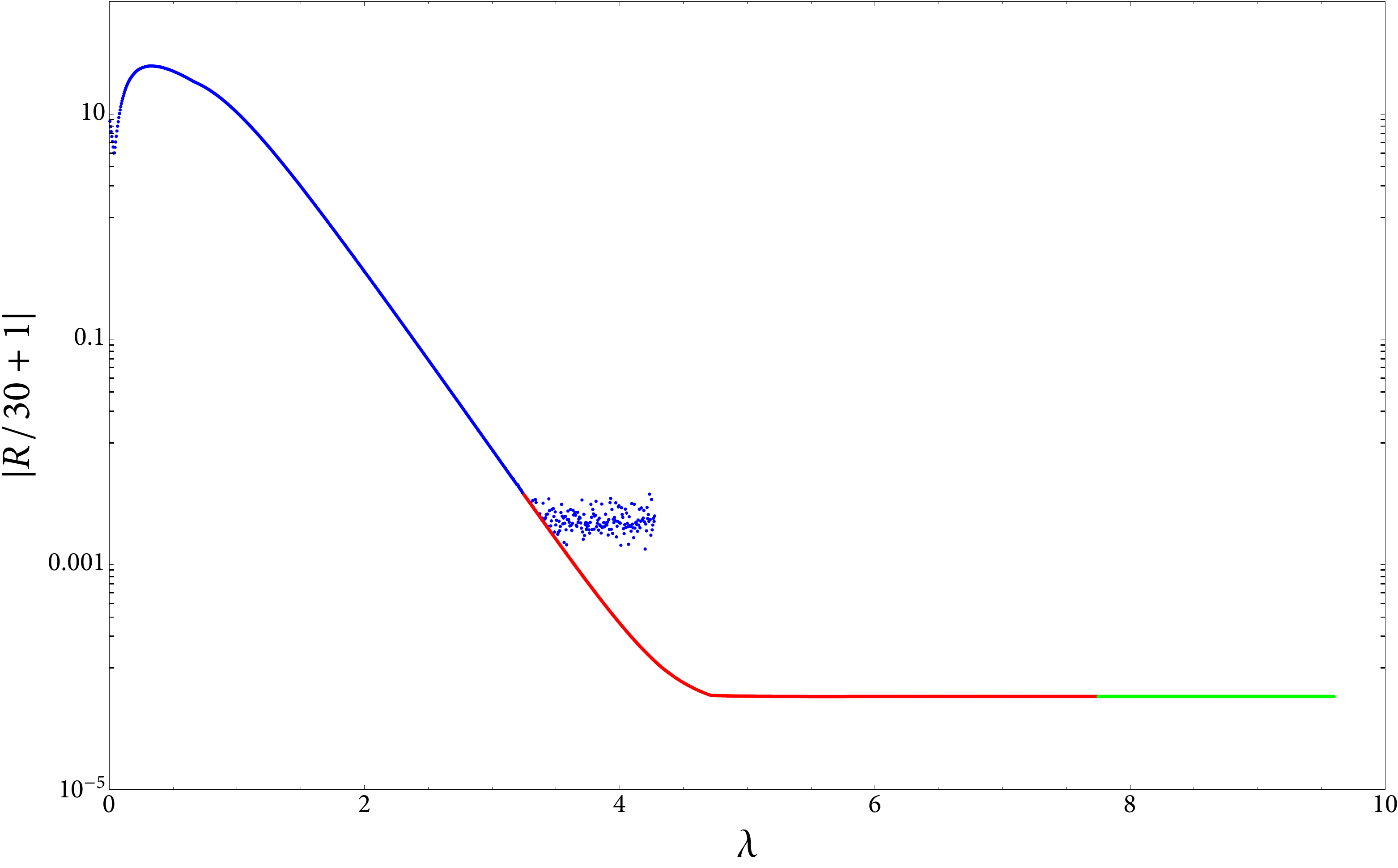}
 \\
 \includegraphics[scale=0.5]{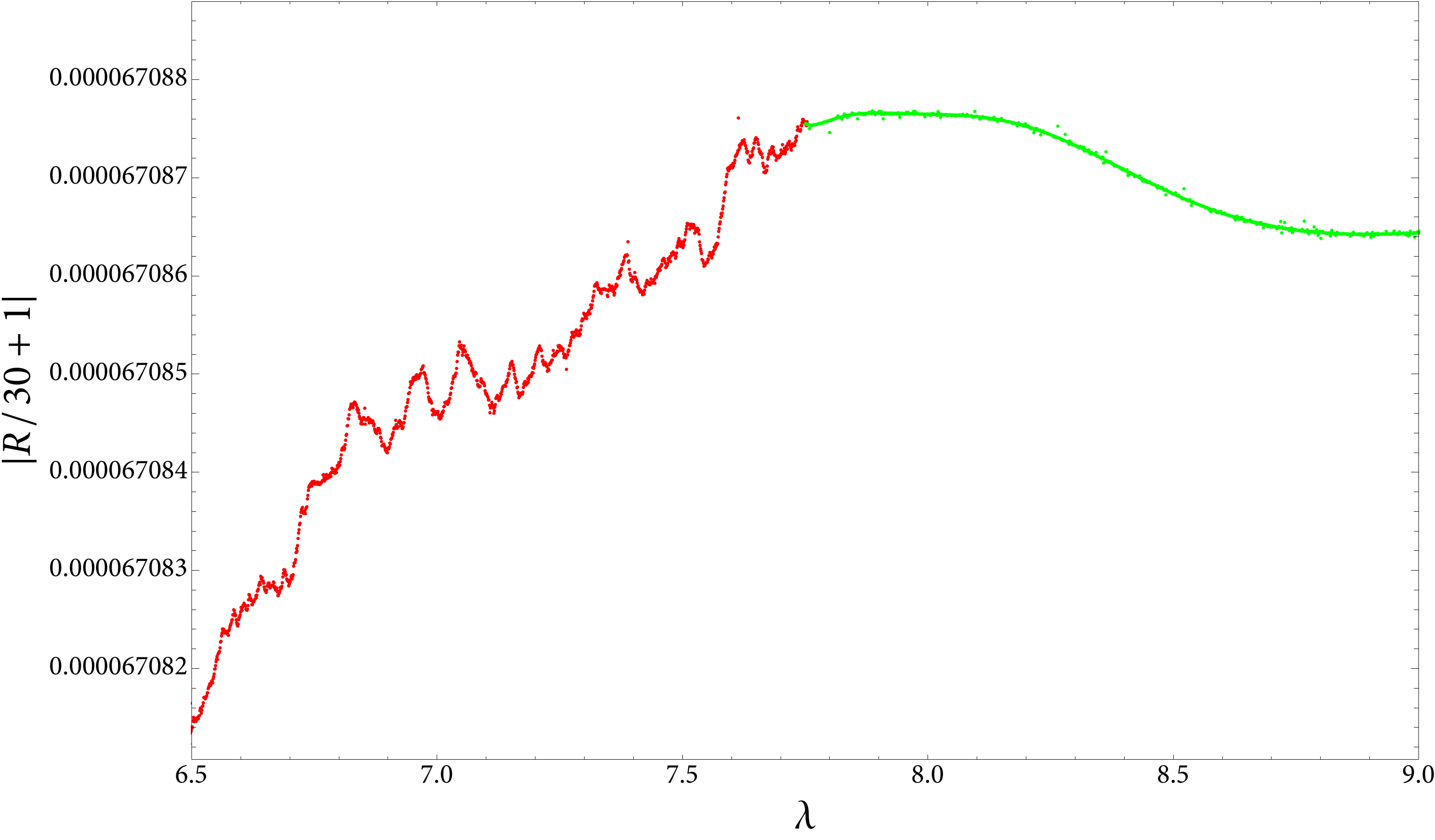} 
 \end{center}
 \caption{Logarithmic plot of $\left|\frac{R}{30}+1\right|_{\textrm{max}}$ as a function of Ricci flow time for a typical run of the simulation. The simulation starts in the blue portion at double precision, then continued using long double precision in the red portion, and finished off using quadruple precision in the green portion.}
 \label{fig:errflow}
 \end{figure}

We should also comment on the behaviour of these quantities as we evolve the Ricci flow. In Fig. \ref{fig:errflow} we plot the maximum of the scalar $\left| \frac{R}{30} + 1 \right|$ for a typical evolution as a function of Ricci flow time. Here we can clearly see the flow stabilising as we approach the fixed point. The figure also shows how insufficient precision causes the flow to oscillate before the fixed point is reached. In all cases, continuing the flow using higher-precision floating point arithmetic allows the flow to progress further. The corresponding plot for $|\phi|_{\textrm{max}}$ exhibits the same behaviour.
 
\pagebreak

\section{Linearised gravity in the higher dimensional RS2 model}
\label{sec:linear}  

In this appendix we sketch the derivation of the linearised gravitational field in the RS2 braneworld model in an arbitrary number of spacetime dimensions. Following the reasoning of \cite{deHaro:2000wj,Figueras:2011va}, the correction to the standard linearised gravity on the brane allows us to read off the fall-off of the stress tensor in the AdS/CFT solution. In particular, for $d$ boundary dimensions we find that  the energy density of the dual field theory, according to the notation in \S\ref{sec:results}, behaves like $1/R^{2d-3}$ in the asymptotic region. For the $d=5$, this gives the $1/R^7$ behaviour, in agreement with our data. 

We closely follow \cite{Garriga:1999yh} (see \cite{Gregory:2008rf} for a review), who studied the weak gravitational field created by an isolated matter source on a 4-dimensional brane. Here we will generalise this result to an arbitrary number of dimensions, so we take $g_{ab}$ to be the metric of a $d$-dimensional brane embedded in a $(d+1)$-dimensional AdS space
\begin{equation}
g_{ab} = \dd{y}^2 + a(y)^2 \, \eta_{\mu\nu} \dd{x}^\mu \dd{x}^{\nu} ,
\end{equation}
where $a(y) = \ee^{-\left| y \right| / \ell}$. The equations of motion for the metric perturbation $g_{ab} \mapsto g_{ab} + h_{ab}$ are
\begin{equation}
 - \frac{1}{2} \left[ a(y)^{-2} \waveop_{(d)} h_{\mu\nu} + a(y)^{-(d-2)} \partial_y \mathopen{} \left( \mathclose{} a(y)^d \partial_y \mathopen{} \left( \mathclose{} a(y)^{-2} h_{\mu\nu} \right) \right) \right]=0\,,
\end{equation}
where we have taken $h_{ab}$ to satisfy the Randall--Sundrum gauge conditions
\begin{equation}
h_{55} = 0 = h_{\mu 5} \quad , \quad h_{\mu \phantom{\nu} ;\nu}^{\phantom{\mu} \nu} = 0 \quad , \quad h^{\mu}_{\phantom{\mu} \mu} = 0 .
\end{equation}

We look for normal modes $h_{\mu\nu} = \psi(y) \ee^{i k^\rho \left(x-x'\right)_\rho}$ with $k^\rho k_\rho = -m^2$. Imposing the Israel junction condition on the brane (see \cite{Garriga:1999yh} for details) and then expanding out the operator, we arrive at an equation for $\psi(y)$:
\begin{equation}
\left[ a(y)^{-2} m^2 + \partial_y \partial_y - \left( d - 4 \right) \ell^{-1} \partial_y - 2 \left( d - 2 \right) \ell^{-2} \right] \psi(y) = -4 \ell^{-1} \delta(y) \psi(y) .
\label{eq:eom}
\end{equation}
Note that in $d = 4$, the coefficient of the first derivative term vanishes, and this expression reduces to the one presented in \cite{Garriga:1999yh}. The general solution to this ODE is given in terms of exponentially-modulated Bessel functions. As the equation is second-order, there are two coefficients to be fixed for each $m$: an overall multiplicative constant which we denote $C_m$ and a relative coefficient which we denote $k_m$. We will now show that latter can be fixed by the aforementioned junction condition while the former can be fixed by an off-brane boundary condition.

When $d$ is odd, the most convenient form of the general solution is
\begin{equation}
\psi_m(y) = C_m \, a(y)^{-(d-4)/2} \left[ J_{-d/2} \mathopen{}\left(\mathclose{} m \ell / a(y) \right) + k_m J_{d/2} \mathopen{}\left(\mathclose{} m \ell / a(y) \right) \right].
\label{eq:psiodd}
\end{equation}
The value of $k_m$ is fixed by the presence of the $\delta$-function. To see this, we integrate (\ref{eq:eom}) over a small neighbourhood $[-\varepsilon,\varepsilon]$ around zero then take $\varepsilon \rightarrow 0$. Since we want $\psi(y)$ to be even under $y \mapsto -y$, we must require $\psi(y)$ to be continuous at $y=0$. Therefore, the $\delta$-function imposes the following \emph{jump condition} on the first derivative of $\psi(y)$: 
\begin{equation}
\psi'\mathopen{}\left(\mathclose{} 0^+ \right) - \psi'\mathopen{}\left(\mathclose{} 0^- \right) = - 4 \ell^{-1} \psi(0) .
\end{equation}
Equation (\ref{eq:psiodd}) then implies that
\begin{equation}
2 m \left( J'_{-d/2}\mathopen{}\left(\mathclose{} m \ell \right) + k_m J'_{d/2}\mathopen{}\left(\mathclose{} m \ell \right) \right) = d\, \ell^{-1} \left( J_{-d/2}\mathopen{}\left(\mathclose{} m \ell \right) + k_m J_{d/2}\mathopen{}\left(\mathclose{} m \ell \right) \vphantom{Y'_{d/2}} \right) ,
\end{equation}
and thus
\begin{equation}
k_m = - \left( \frac{2 m J'_{-d/2}\mathopen{}\left(\mathclose{} m \ell \right) + d \ell^{-1} J_{-d/2}\mathopen{}\left(\mathclose{} m \ell \right)}{2 m J'_{d/2}\mathopen{}\left(\mathclose{} m \ell \right) + d \ell^{-1} J_{d/2}\mathopen{}\left(\mathclose{} m \ell \right)} \right) .
\label{eq:kodd}
\end{equation}
In the following we will also need the expansion of $J_\nu(x)$ about zero. For completeness we record it below:
\begin{equation}
J_{\nu}(x) \sim \frac{x^\nu}{2^\nu \Gamma\mathopen{}\left(\mathclose{} 1+\nu \right)} - \frac{x^{\nu+ 2}}{2^{\nu + 2} \Gamma\mathopen{}\left(\mathclose{}  2 + \nu \right)} + \ldots \qquad \textrm{as } x \rightarrow 0.
\end{equation}
Inserting this into (\ref{eq:kodd}) gives
\begin{align}
k_m &\sim \frac{\Gamma\mathopen{}\left(\mathclose{} d/2 \right)}{\Gamma\mathopen{}\left(\mathclose{} 2 - d/2 \right)}  \left( \frac{2}{m \ell} \right)^{d-2} \qquad \textrm{as } m \ell \rightarrow 0.
\end{align}

When $d$ is even, the most convenient form of the general solution is
\begin{equation}
\psi_m(y) = C_m \, a(y)^{-(d-4)/2} \left[ Y_{d/2} \mathopen{}\left(\mathclose{} m \ell / a(y) \right) + k_m J_{d/2} \mathopen{}\left(\mathclose{} m \ell / a(y) \right) \right].
\label{eq:psieven}
\end{equation}
The jump condition now reads
\begin{equation}
2 m \left( Y'_{d/2}\mathopen{}\left(\mathclose{} m \ell \right) + k_m J'_{d/2}\mathopen{}\left(\mathclose{} m \ell \right) \right) = d \ell^{-1} \left( Y_{d/2}\mathopen{}\left(\mathclose{} m \ell \right) + k_m J_{d/2}\mathopen{}\left(\mathclose{} m \ell \right) \vphantom{Y'_{d/2}} \right) ,
\end{equation}
and thus
\begin{equation}
k_m = - \left( \frac{2 m Y'_{d/2}\mathopen{}\left(\mathclose{} m \ell \right) + d \ell^{-1} Y_{d/2}\mathopen{}\left(\mathclose{} m \ell \right)}{2 m J'_{d/2}\mathopen{}\left(\mathclose{} m \ell \right) + d \ell^{-1} J_{d/2}\mathopen{}\left(\mathclose{} m \ell \right)} \right) .
\label{eq:keven}
\end{equation}
We now need the expansion of $Y_n$ about zero for $n \in \mathbb{N}$:
\begin{equation}
Y_n(x) \sim -\frac{2^n \left( n - 1 \right)!}{\pi x^n} - \frac{2^{n-2} \left( n - 2 \right)!}{\pi x^{n-2}} - \ldots \qquad \textrm{as } x \rightarrow 0 .
\end{equation}
Therefore, in this case $k_m$ is given asymptotically by
\begin{align}
k_m &\sim \frac{\left( d/2 - 1 \right)! \left(d/2 - 2 \right)!}{\pi} \left( \frac{2}{m \ell} \right)^{d-2} \qquad \textrm{as } m \ell \rightarrow 0.
\end{align}
We can summarise these results to apply in any number of dimensions $d$ by writing $k_m \sim \frac{\kappa_{(d)}}{(m \ell)^{d-2}}$, where $\kappa_{(d)}$ is a constant which depends only on $d$.

The overall constant $C_m$ is determined by requiring that the eigenmodes are finite as $m\to 0$ for any value of $|y|$. We can therefore consider the behaviour of the eigenmodes as we move far away from the brane, taking $|y|$ to be sufficiently large. The large argument asymptotics of the Bessel functions are given by
\begin{align}
J_{\nu}\mathopen{}\left(\mathclose{} x \right) \sim \sqrt{\frac{2}{\pi x}} \cos\mathopen{}\left(\mathclose{} x - \frac{(2 \nu + 1) \pi}{4} \right) & \\
Y_{\nu}\mathopen{}\left(\mathclose{} x \right) \sim \sqrt{\frac{2}{\pi x}} \sin\mathopen{}\left(\mathclose{} x - \frac{(2 \nu + 1) \pi}{4} \right) & \qquad \textrm{as } x \rightarrow \infty .
\end{align} 
For convenience, let us write $\hat{y} \equiv a(y)^{-1} = \ee^{|y| / \ell}$. Then, for large $m\ell\hat y$, the eigenmodes behave like
\begin{equation}
\psi_m\mathopen{}\left(\hat{y}\right) \sim C_m \sqrt{ \frac{2 \hat{y}^{d-5}}{\pi m \ell} } \left[ \sin\mathopen{}\left(\mathclose{} m \ell \hat{y} - \frac{(d + 1) \pi}{4} \right) + \frac{\kappa_{(d)}}{(m \ell)^{d-2}} \cos\mathopen{}\left(\mathclose{} m \ell \hat{y} - \frac{(d + 1) \pi}{4} \right) \right] .
\end{equation}
Requiring that the amplitude of these modes neither vanishes nor blows up as $m \rightarrow 0$, in the limit of $m\ell\hat y\to\infty$, fixes the small-$m$ dependence of $C_m$  to be (neglecting any numerical factors)
\begin{equation}
C_m \sim (m \ell)^{d-3/2} \qquad \textrm{as } m \ell \rightarrow 0 .
\end{equation}
Finally, since $C_m$ is a constant independent of $y$, the above dependence must also hold on the brane $y=0$. We can therefore read off the small-$m$ dependence of $\psi_m$ as
\begin{equation}
\psi_m \sim (m \ell)^{(d-3)/2} .
\label{eqn:psiasym}
\end{equation}

The gravitational potential is basically given by the Green's function for (\ref{eq:eom}), which consists of a superposition of all the eigenfunctions $\psi_m$ that we have just derived. We are only interested in the gravitational field on the brane itself in the far field region, henceforth we set $y=0$ and we do not keep track of the overall numerical coefficients.   Note here that there is a continuum of KK eigenmodes as well as a discrete zero mode, thus
\begin{equation}
\left. G(x,x') \right|_{y=0} = - \int \frac{\dd^d k}{(2\pi)^d} \ee^{i k_\mu (x-x')^\mu} \left( \frac{\ell^{-1}}{\left| \mathbf{k} \right|^2 - \omega^2} + \int_0^\infty \dd m \frac{\psi_m(0)^2}{\left| \mathbf{k} \right|^2 + m^2 - \omega^2} \right) .
\end{equation}
Since we also only interested in the stationary state, we can integrate out the $t'$-dependence, leaving just
\begin{equation}
\left. G(\mathbf{x}, \mathbf{x}') \right|_{y=0} = - \int \frac{\dd^{d-1}\mathbf{k}}{(2\pi)^d} \ee^{i \mathbf{k} \cdot \left( \mathbf{x}-\mathbf{x'} \right)} \left( \frac{\ell^{-1}}{\left| \mathbf{k} \right|^2} + \int_0^\infty \dd m \frac{\psi_m(0)^2}{\left| \mathbf{k} \right|^2 + m^2} \right) .
\end{equation}
The first term gives rise to the usual $1/r^{d-3}$ potential for standard gravity on the brane, whilst the second term is responsible for the correction due to the KK modes and this is the relevant term for us. Thus, we define
\begin{equation}
V_{KK} = - \int \frac{\dd^{d-1}\mathbf{k}}{(2\pi)^d} \ee^{i \mathbf{k} \cdot \left( \mathbf{x}-\mathbf{x'} \right)} \left( \int_0^\infty \dd m \frac{\psi_m(0)^2}{\left| \mathbf{k} \right|^2 + m^2} \right) \,.
\end{equation}
Changing the order of the integration and doing first the integral over the momenta $\mathbf k$, gives rise to a Yukawa-type potential
\begin{equation}
V_{KK} = \frac{1}{r^{d-3}} \, \frac{\Gamma(d/2-1/2)}{2 (d-3) \pi^{(d-1)/2} } \int_0^{\infty} \dd m\,  F_{(d-3)/2}(mr) \psi_m(0)^2\,,
\label{eqn:vkktmp}
\end{equation}
where we have defined $F_\nu(\xi)\equiv \frac{1}{2^{\nu-1}\Gamma(\nu)}\xi^\nu K_{\nu}(\xi)$, and $K_\nu$ is the modified Bessel function of the second kind. Indeed, for $d=4$ we have precisely $F_{1/2}(mr) = \ee^{-mr}$. For a general $d$, $F_{(d-3)/2}$ generalises this exponential screening, as its asymptotic behaviour is given by
\begin{equation}
F_{(d-3)/2}(mr) \sim \frac{\sqrt{\pi}\left(mr\right)^{d/2-2} \ee^{-mr}}{2^{d/2-2} \Gamma(d/2-3/2)} \left[ 1 + \mathcal{O}\mathopen{}\left(\mathclose{} (mr)^{-1} \right) \right]\,.
\label{eqn:yukawa}
\end{equation}

We extract the leading order correction to the linearised gravitational potential by integrating over $m$. As in \cite{Randall:1999vf,Garriga:1999yh}, we note that in the large-$r$ limit the integral is dominated by contribution from the small-$m$ modes. Using \eqref{eqn:psiasym} in \eqref{eqn:vkktmp} yields
\begin{equation}
V_{KK} \sim \frac{1}{r^{2d-5}} \qquad \left( r \gg 1 \right),
\end{equation}
so that the full gravitational potential on the brane is
\begin{equation}
V(r)\sim \frac{G_N\,M}{r^{d-3}}\left(1+\frac{\alpha\,\ell^{d-2}}{r^{d-2}}\right)
\label{eqn:gravpotential}
\end{equation}
where $\alpha$ is some non-vanishing (dimensional dependent) numerical factor and $G_N$ is Newton's constant on the brane. Then, the metric perturbation on the brane will be 
\begin{equation}
h_{00} \sim V(r)\,.
\label{eqn:perturbation}
\end{equation}
The other components of the metric perturbation can be shown to exhibit a similar behaviour in the far field region. 

Following \cite{deHaro:2000wj,Figueras:2011gd}, considering the limit in which the brane is close to the boundary of AdS, one can use the correction to the Einstein tensor on the brane induced by \eqref{eqn:perturbation} to read off the components of the stress tensor in the far field regime:
\begin{equation}
\delta G_{\mu\nu}=16\,\pi\,G_N \langle T_{\mu\nu}^{CFT}\rangle\,.
\end{equation}
Note that only the term proportional to $\alpha$ in \eqref{eqn:gravpotential} contributes to the left hand side of this equation, which implies
\begin{equation}
 \langle T_{\mu\nu}^{CFT}\rangle \sim \frac{1}{r^{2d-3}}\,.
\end{equation}
For $d=5$, this is precisely the same behaviour that our data exhibits.

 
 \newpage
  
\bibliographystyle{utphys}
\bibliography{RotatingBdry}

\end{document}